# Inoculating solid-state homogeneous precipitation by impurity atoms through a spinodal decomposition like pathway

Shiwei Pan [1,3], Chunan Li [1,3], Hanne-Sofie Søreide [1,3], Dongdong Zhao [1], Constantinos Hatzoglou [1], Feng Qian [1] ✉, Long-Qing Chen[2], Yanjun Li [1] ✉

[1] Department of Materials Science and Engineering, NTNU, Trondheim, Norway

[2] School of Materials Science and Engineering, Penn State University, USA

[3] These authors contributed equally.

✉ e-mail: qianfeng1101@163.com, yanjun.li@ntnu.no

## Abstract

**Solid-state homogeneous precipitation of nano-sized precipitates is one of the most effective processes to strengthen metal alloys, where the final density and size distribution of precipitates are largely controlled by the precipitation kinetics. Here, we report a strategy to inoculate the homogeneous precipitation of coherent precipitates to enhance the precipitation strengthening. Using the technologically important dilute Al-Zr alloys as an example, we demonstrate that an addition of a trace level of economical and readily available, non-$L1_2$ phase forming impurity atoms, X (X= Sn, Sb, Bi or Cd) and Si, can significantly enhance the diffusivity of Zr atoms and overturn the precipitation of $L1_2$-structured $Al_3Zr$ nanoparticles from the classical homogeneous nucleation and growth pathway into a nonclassical nucleation pathway: $Al_3Zr$ forms through the spontaneous formation of nano-scale local concentration fluctuations of Zr atoms on Zr-X(-Si)-vacancy clusters followed by a continuous increase of the concentration and chemical short-range ordering (CSRO). Such an impurity atoms induced heterogeneous nucleation based on a "spinodal decomposition like" mechanism dramatically accelerates the precipitation kinetics, leading to an order of magnitude higher number density of precipitates and a record high hardening efficiency of solute Zr atoms. By formulating the generalized selection principles for inoculating impurity elements, this inoculation strategy should be extendable to a broader range of materials to further explore the precipitation strengthening potentials.**

## Main content

Solid-state precipitation has been the often utilized and most effective process to strengthen a wide variety of alloys, and the degree of strengthening is largely dependent on the density and size distribution of precipitates. It has been commonly believed that homogeneous precipitation in solids takes place through two well-known conventional mechanisms. One is through the classical nucleation and growth pathway, mostly occurring in metastable, and typically dilute, solid solutions[1]. It starts with the formation of stable nuclei which have the same crystal structure, chemical order and chemistry as the precipitate phase described by the corresponding phase diagram [2,3]. The second is spinodal decomposition of unstable solid solutions where the new phase forms via spontaneous formation and gradual growth of non-localized spatially extended coherent composition fluctuations and/or long-range ordering [4]. Precipitation by spinodal decomposition often occurs in concentrated alloys with a miscibility gap. There have also been recent reports of nonclassical nucleation pathways in which the formation of a new phase is through intermediate states, which are different from either the parent or the product phases. Such non-classical nucleation mechanisms are more common in crystallization of nanoparticles from liquid or gas solutions, where the bulk crystal phases can be formed by aggregation of nano-sized precursors including amorphous solids, atom complexes, poorly crystalline particles, and nanocrystals [5–9].

The largest energy barrier for nucleation of precipitates in solid through the classical nucleation and growth pathway stems from the interfacial energy and strain energy between the nucleus and matrix. A prominent example is the homogeneous precipitation of $L1_2$-structured $Al_3Zr$ in technologically important Al-Zr based alloys, which is known through the classical nucleation and growth mechanism [10]. Due to the super low solubility (max 0.083 at. %) and extremely low diffusivity of Zr in Al solid solution, the driving force for nucleation of $Al_3Zr$ in Al-Zr binary alloys is low and therefore the precipitation kinetics is very sluggish and the achievable precipitation hardening is rather limited. Inoculation by introducing high density potent heterogeneous nucleation sites is an effective method to reduce the nucleation energy barrier, enhancing the nucleation processes. This strategy has been widely applied in metallurgy industry to refine grains and crystals in metal alloys during solidification [11] and in synthesis of diamond particles by introducing carbon atom clusters as protonuclei [12] or pre-assembled colloidal clusters [13]. However, for solid state precipitation of nano-sized coherent particles like $Al_3Zr$, such an inoculation method is challenging due to the difficulty to generate suitable nano-sized nucleation sites. So far, this has been achieved by adding rather expensive Sc or Er in Al-Zr alloys to form high density $L1_2$-structured $Al_3Sc$ [14–17] or $Al_3Er$ [18] as seeding precipitates for $Al_3Zr$, accelerating the subsequent nucleation of $Al_3Zr$ precipitates, forming core-shell structured $Al_3(Sc,Zr)$ or $Al_3(Er,Zr)$. However, the extremely high price of Sc and low solubility of Er in Al solid solution have greatly limited their industrial applications. In this work, we are showing a new strategy to inoculate the precipitation of $Al_3Zr$ nanoprecipitates in Al-Zr alloys by adding non-$L1_2$ phase forming economical impurity elements to change the precipitation mechanism from the classical nucleation and growth pathway into a “spinodal like” heterogeneous nucleation pathway, and therefore dramatically enhance the precipitation kinetics.

We fabricated an Al-0.08Zr and an Al-0.08Zr-0.1Si-0.02Sn (in at.%) alloy by conventional metal mold casting method [19]. Fig. 1a shows that an addition of trace levels of Sn and Si has dramatically enhanced the age-hardening kinetics of Al-Zr alloys, achieving higher peak hardness within a much shorter time. At the peak aged state, the yield strength of the Al-Zr-Sn-Si alloy is about 2.2 time as that of the Al-Zr binary alloy (Fig. 1b and Extended Data Table 2). Figs. 1d and e compare the distribution of $Al_3Zr$ in dark-field transmission electron microscopy (TEM) images in the two alloys after peak aging (more images are shown in supplementary Fig. S1 and S2). The insets in the pictures show the electron diffraction patterns of the precipitates, confirming they own $L1_2$ structure. Quantitative analyses show that the number density of precipitates in Al-Zr-Sn-Si is an order of magnitude higher ($4.5\times10^{22}$ v.s. $5.4\times10^{21}$ $m^{-3}$) while the average size (3.4 v.s. 4.8 nm in radius) is much smaller than Al-Zr alloy. The faster precipitation kinetics of $Al_3Zr$ in Al-Zr-Sn-Si alloy is also clearly indicated by the faster increase of electrical conductivity (Fig. S3).

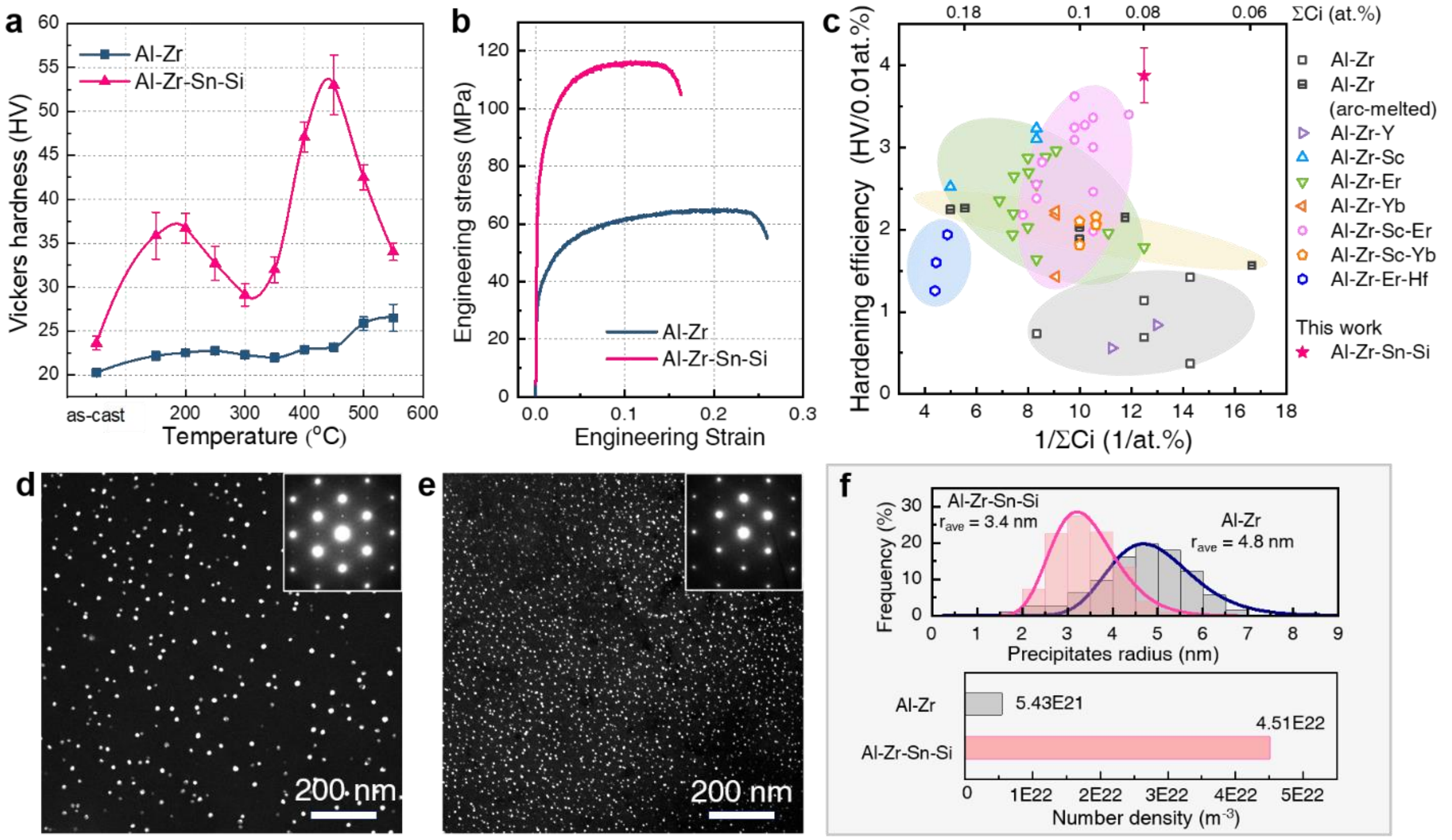

**Fig. 1. The promoted precipitation kinetics and hardening effect by inoculation. a,** Microhardness evolution of Al-Zr-Sn-Si and Al-Zr alloys during multi-steps isochronal aging from 25 to 550 °C. **b,** Engineering stress strain curves at peak-aged states in **a**. **c,** Precipitation hardening efficiency (hardness increase normalized by 0.01 at. % of all $L1_2$-forming element, $\sum Ci$, of peak-aged Al-Zr-Sn-Si in comparison to Al-Zr alloys reported in literatures [19]. **d** and **e,** Dark-field TEM images showing the $Al_3Zr$ precipitates under the peak-aged condition. **f,** Size distributions and number densities of $Al_3Zr$ precipitates.

Fig. 1c compares the hardening efficiency $E$, maximum hardness increment by per 0.01 at. % of $L1_2$-forming elements (sum of Zr, Sc, Er and Yb), of the present Al-Zr-Sn-Si alloy to other Al-Zr based alloys in literatures, showing that inoculation by impurity Sn and Si has increased $E$ value of Al-Zr alloy from ~0.5 to 3.8, which is even superior to the Al-Zr-Sc-Er alloys (~3.5).

**Non-classical nucleation of $Al_3Zr$**

More interestingly, the Al-Zr-Sn-Si alloy shows two distinctive hardness peaks during isochronal heating with the first one appearing at 200 °C, and the second one at 450 °C, suggesting other type of precipitates than $Al_3Zr$ may have precipitated in the early stage of aging. To explore the underlying mechanism, we conducted a series of pre-aging treatments at 200 and 300 °C, which are known too low to cause $Al_3Zr$ precipitation within relatively short time. It shows that pre-aging treatments at both temperatures can further speed up the precipitation kinetics and increase the achievable peak hardness of the alloy during the subsequent 400 °C isothermal aging (Extended Data Fig. 1). Strikingly, Al-Zr-Sn-Si alloy pre-aged at 300 °C for 8h shows a continuous increase of both hardness and electrical conductivity already after 10 min aging at 400 °C. In contrast, it takes 1h and 24h for un-preaged Al-Zr-Sn-Si alloy and the reference Al-Zr alloy, respectively, to achieve any detectable hardness increase. It indicates that the incubation time required to reach steady state nucleation of $Al_3Zr$ has been substantially shortened by 300 °C pre-aging. Since the same pre-aging treatments have negligible effect on the aging of reference Al-Zr binary alloy at 400 °C (Fig. S4), the addition of Sn and Si has clearly catalyzed the precipitation of $Al_3Zr$.

Based on the classical nucleation theory [20,21], we have calculated the critical nucleation energy barrier $\Delta G^*$ and incubation time $\tau$ for precipitation of $Al_3Zr$ in binary Al-0.08Zr alloy at different temperatures (detailed calculation is shown in Extended data and Table S2). At 400 °C, the $\Delta G^*$ value is calculated as 37 $k_B$T ($k_B$ stands for Boltzmann constant and T is temperature in K), showing that thermal fluctuation is sufficient to trigger the formation of precipitate nuclei [3,22]. However, due to the extremely low diffusivity of Zr, the incubation time $\tau$ needed to reach steady state nucleation is 23.6 ~ 55.6 h according to different equations [20,21], which is very close to the experimental result (24 h), confirming the classical nucleation behavior in binary Al-Zr alloy. If we assume the solubility and diffusivity of Zr are not changed by the addition of Sn and Si impurities, a reduction of incubation time from 24h to 1h implies a reduction of interfacial energy between $Al_3Zr$ and Al matrix from 0.1 $J/m^2$ [10,23] to 0.004 $J/m^2$ in and correspondingly a super low $\Delta G^*$ value of $2.37\times10^{-3}$ $k_BT$ in Al-Zr-Sn-Si alloy, which is unlikely. For the 300 °C pre-aged Al-Zr-Sn-Si alloy, the measured incubation time as low as 10 minutes implies an even lower interfacial energy.

We further characterized the microstructures in Al-Zr-Sn-Si alloy generated during the pre-aging treatments by atom probe tomography (APT) and high-angle annular dark-field scanning transmission electron microscopy (HAADF-STEM). After 24h pre-aging at 200 °C, only Sn-rich atom clusters ($N_v$ = $9.9\times10^{21}$ $m^{-3}$) with an average radius of ~2 nm can be detected in the reconstructed APT volume (Fig. 2a). The formation of such Sn-rich clusters is supposed to be responsible for the hardness increase during pre-aging and the first hardness peak shown in Fig. 1a. More interestingly, a remarkably higher concentration of Zr and Si atoms than their nominal concentrations can also be detected in the Sn-rich clusters (Fig. 2b), implying a strong clustering tendency of Zr and Si atoms with Sn atoms. In the Al matrix, some atom columns showing higher contrast can be observed in high-resolution HAADF-STEM images (Fig. S5 b-g) implying an enrichment of Sn and/or Zr solute atoms. By excluding Sn-rich atom clusters from the APT needle sample, normalized partial radial distribution function (PRDF) profile of Zr around Sn atoms shows significantly higher values than 1 (representing random distribution) within 1 nm range (Extended Data Fig.2), indicating nano-sized local chemical concentration fluctuations in the form of (Zr, Sn)-rich atom complexes have formed in the Al matrix.

After 8h pre-aging at 300 $^{o}$C, Sn-rich nanoparticles containing Zr can also be observed occasionally by TEM (Fig. S5), but with a much larger size and lower density, indicating a significant coarsening of Sn-rich particles. This explains the hardness decrease at around 300 $^{o}$C in Fig. 1a. In APT samples as shown in Fig. 2c, Sn-rich clusters cannot be observed anymore; instead, a high density of Zr-rich atom clusters can be detected by computing 1.0 at.% Zr iso-concentration surfaces. Based on the iso-position method, the number density and average radius of Zr-rich clusters are determined as $5.2\times10^{22}$ m$^{-3}$ and 0.57 nm, respectively. Proxigram studies reveal that the clusters show a gradual increase of Zr concentration from the edge to the center, with the maximum concentration of Zr in the centers of large clusters measured as ~ 9 at.% (Fig. 2d), confirming they are not $Al_3Zr$ but Zr-rich atom clusters. Inside these clusters, an enrichment of Si, and a rather weak enrichment of Sn can also be detected. A comparison of the Zr-Zr and Zr-Sn PRDF curves between 200 $^{o}$C and 300 $^{o}$C pre-aged samples (Fig. 2e) show that the amplitude of local concentration fluctuations of Zr atoms is higher while the Sn concentration in the clusters is lower in the latter sample. The stronger clustering of Zr atoms during 300 $^{o}$C pre-aging can also be distinguished by the nearest neighbor distribution profiles shown in Fig. S7. Although it is difficult to precisely determine the atomic positions of Zr, Sn and Si atoms in the clusters by APT, the high concentration of Zr-Zr locating at the 1NN and 2NN positions suggests that a chemical short-range ordering (SRO) of Zr atoms may have occurred within the clusters during pre-aging.

Zr-rich clusters formed during 300 $^{o}$C pre-aging are further confirmed by HAADF-STEM images in Fig. 2f-g, where a high density ($6.5\times10^{22}$ m$^{-3}$) of bright clusters with an average radius of ~0.48 nm (Extended Data Fig. 4d) can be observed. Although the atomic-resolution images of the clusters and the corresponding fast Fourier transform (FFT) patterns (Fig. 2f, g and Extended Data Fig. 4) do not show the long range $L1_2$ ordering of $Al_3Zr$, the alternating intensity of atom columns in the clusters confirms the existence of local CSRO of Zr atoms (Fig. 2h). With the high Zr concentration and local CSRO, the clusters are expected to easily evolve into $L1_2$-structured $Al_3Zr$ during subsequent 400 °C aging, which is responsible for the significantly faster precipitation kinetics caused by 300°C pre-aging.

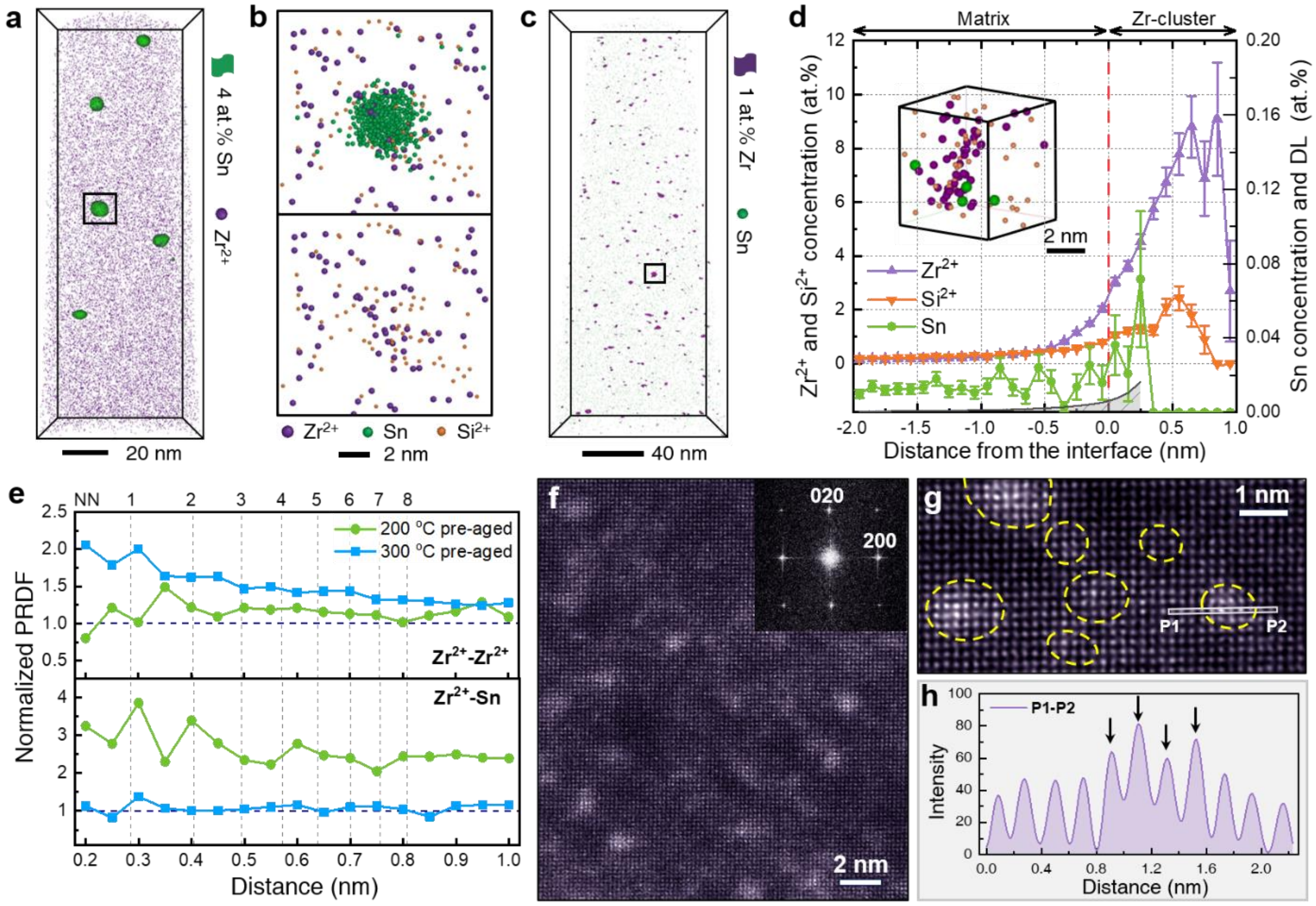

**Fig. 2. Formation of short-range ordered Zr-rich atom complexes and clusters in Al-Zr-Sn-Si alloy after pre-aging. a,** Reconstructed APT volume of 200 °C/ 24 h specimen, illustrating the presence of Sn-rich atom clusters (4 at.% Sn iso-surfaces) and the distribution of $Zr^{2+}$ ions (purple). **b,** Distribution of solute atoms in/around a typical Sn-rich cluster framed by a black box in **a**. **c,** Reconstructed APT volume of 300 °C/ 8h specimen, showing Zr-rich atom clusters distinguished by 1% Zr iso-surface (colored in purple) containing nine or more solute atoms. **d,** Proximity histogram of Zr-rich clusters **c** and the inset is one typical Zr-rich cluster. **e,** Normalized partial radial distribution function (PRDF) of $Zr^{2+}$ ($Zr^{2+}$-$Zr^{2+}$, $Zr^{2+}$-Sn) within the reconstructed tips in **a** and **c**, respectively. **f-g,** Atomic-resolution HAADF-STEM images of 300 °C/8h specimen, with Zr-rich clusters of higher contrast circled by yellow dashed lines in **g**. **h,** Intensity profile of atom columns along P1-P2 line across one cluster in **g**, showing a subtle periodical intensities variation of 4 columns (indicated by arrows).

We estimated the growth velocity of atom clusters during 300 $^{o}$C pre-aging by treating the clusters as spherical precipitates. It is calculated that within 8h, the radius of clusters with uniform Zr concentrations of $C_p = 1\ at.\%$ and $5\ at.\%$ can reach a maximum value of 0.18 and 0.07 nm, respectively (detailed calculation and results are shown in Method section and Fig. S8). They are much smaller than the measured average radius, showing the addition of Sn and Si has substantially enhanced the growth kinetics of Zr-rich clusters, corresponding to an order of magnitude higher diffusivity of Zr.

**Role of Sn**

To distinguish the roles played by Sn and Si, we further studied the age-hardening behavior of a ternary Al-0.08Zr-0.02Sn alloy (Fig. S3) and the influence of pre-aging. As shown by the hardness curve at 400 $^{o}$C in Fig. 3a, the addition of only Sn (0.02 at. %) could reduce the incubation time from 24h to 4h and increase the peak hardness value from 26 to 41.8 HV. A 200 $^{o}$C/24h pre-aging further reduces the incubation time to 2h, achieving a hardness of 44 HV within 48 hours. Same as Al-Zr-Sn-Si alloy, high density nano-sized Sn-rich clusters containing a relatively high content of Zr can also be detected in the pre-aged ternary alloy by APT analyses (Fig.3b, c and Movie. S2). Normalized PRDF results (Fig. 3d) also reveal the strong clustering tendency between Zr and Sn atoms in the ternary alloy. These results demonstrate that Sn atoms have played a key catalyzing role in the formation of the concentration fluctuation of Zr in forms of Sn-Zr rich clusters.

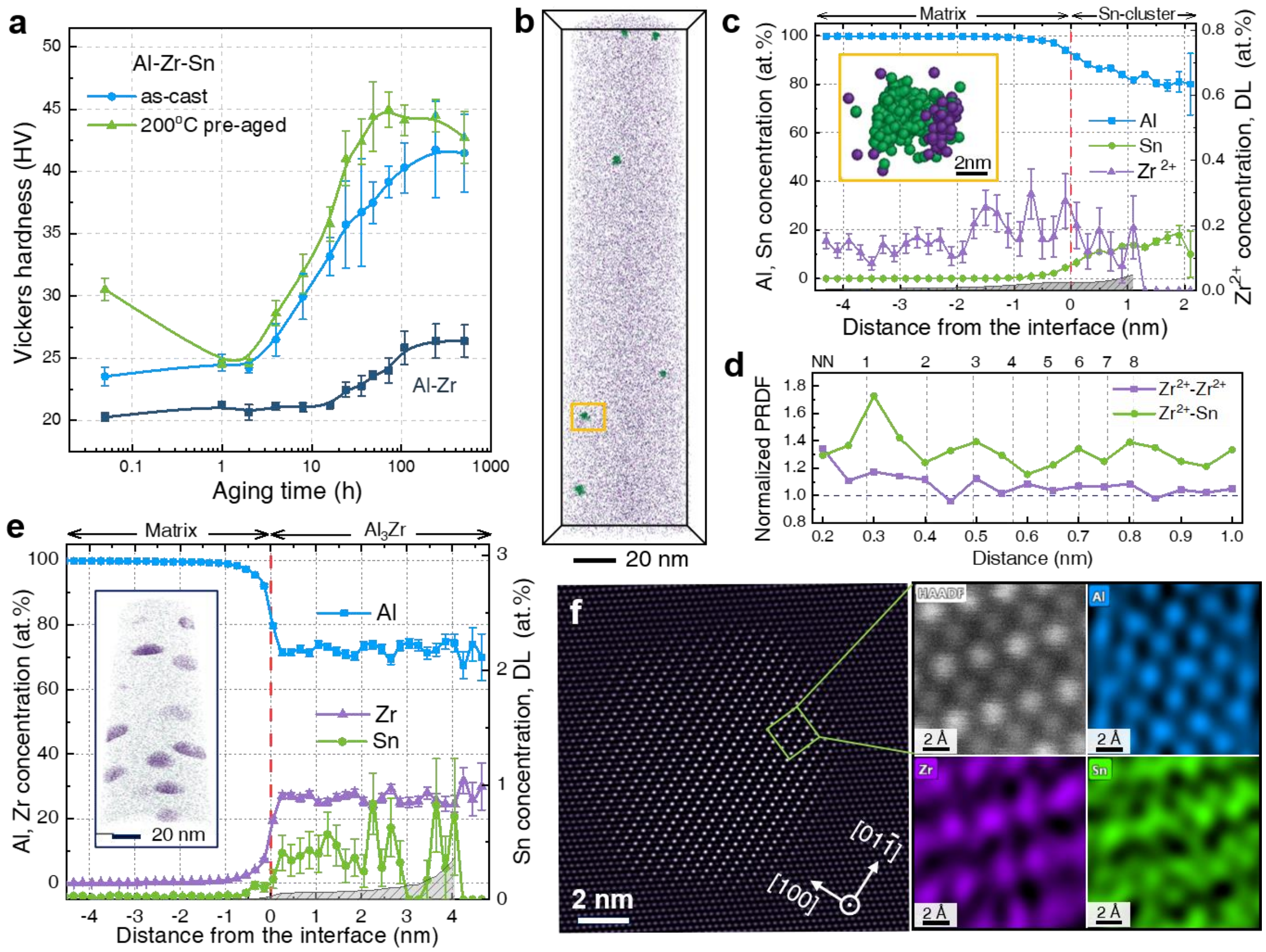

**Fig. 3. Inoculation effect of Sn on ternary Al-Zr-Sn alloy. a,** Evolution of microhardness of the as-cast and 200 °C/24 h pre-aged samples during isothermal aging at 400 °C. **b-d,** APT analysis of the 200 °C/24h sample, showing spatial distribution of Zr (purple) and Sn atoms (green). **c**, Proximity histogram of Sn-rich clusters computed from a 4% Sn isosurface, the inset is a typical Sn-rich cluster framed in **b**. **d**, Normalized PRDF of Zr-Zr and Zr-Sn within the APT tip. **e,** Proximity histogram of $Al_3Zr$ precipitates computed by 3% Zr iso-surface under the peak aged condition (200 °C/24 h + 400 °C/72 h), with an inset of the reconstructed APT tip showing the distribution of precipitates. **f,** HAADF-STEM image of a $Al_3Zr$ precipitate in the peak-aged condition taken along a <011>$_{Al}$ zone axis, and the corresponding atomic-scale EDX mapping of a local region (marked by a green rectangle) of the precipitate.

To prove that the $L1_2$-structured $Al_3Zr$ precipitates have developed from the (Zr, Sn)-rich atom clusters, the chemistry of the $Al_3Zr$ precipitates at the peak-aged state (200 °C /24h+400 °C /48h) of Al-Zr-Sn alloy was studied by APT (Fig.3e). In the proxigram of $Al_3Zr$ precipitates delineated by 3 at.% Zr iso-surfaces in the reconstructed volume, a much higher concentration of Sn (0.4-0.5 at.%) than its nominal content can be identified. The atomic-resolution EDX mapping of a typical $Al_3Zr$ precipitate in HAADF-STEM mode (Fig. 3f) shows a preferential positioning of impurity Sn atoms in mixed Al+Zr atom columns of the $L1_2$ precipitate instead of pure Al atom columns, implying Sn may have replaced some Zr atoms. However, our DFT calculations (Table S3) show that a substitution of either Zr or Al atoms in $Al_3Zr$ by Sn is energetically unfavorable, indicating that the detected impurity Sn atoms are not due to the partitioning of Sn into $Al_3Zr$ precipitates during growth, but stemmed from the (Zr, Sn)-rich clusters formed during the early stage of aging which act as a precursor of $Al_3Zr$. Since it is difficult for vacancy, and therefore Sn, to migrate within $L1_2$-$Al_3Zr$ [17], the inherited Sn atoms were trapped in the precipitate. A slight enrichment of Sn and Si was also observed in $Al_3Zr$ precipitates in the peak-aged quaternary Al-Zr-Sn-Si alloy (Extended Data Fig.3d), which should also be inherited from the (Zr, Sn, Si)-rich clusters formed in the early stage of heat treatment.

**Impurity induced concentration fluctuation and accelerated diffusion of Zr.**

Next, in Fig. 4 and Extended Data Fig. 5, Table 3 and 4, we used DFT calculations to explore the underlying mechanism for the formation of (Zr, Sn, Si)-rich atom clusters by calculating the binding energies of Zr-Va dimer, Zr-Sn-Va and Zr-Si-Va trimers (Va stands for vacancy) with different configurations in Al lattice. As expected, the 1NN Zr-Va dimer shows a repulsive binding energy of -0.21 eV, responsible for the extremely low diffusivity of Zr in Al. However, when a vacancy is bonded with both Zr and Sn atoms, strong attractive binding energies of 0.14 eV (Zr-Va at 1NN) and 0.35 eV (Zr-Va at 2NN) can be reached for some configurations. By forming such low energy Sn-Zr-Va complexes and the further growth of such complexes into larger clusters by absorbing more Zr and Sn atoms, local concentration fluctuations in the solid solution can be easily generated. On the other hand, by forming such Sn-Zr-Va trimers, the chance for a Zr atom to access a vacancy at the 1NN positions in Al lattice is significantly increased, which will accelerate the diffusivity of Zr. In contrast, the Si-Zr-Va trimers show only slightly improved attractive binding energy in comparison to Zr-Va dimers, implying only Si addition exerts a much less influence in catalyzing the Zr concentration fluctuation than Sn. This is consistent with the limited improvement of precipitation hardening in ternary Al-Zr-Si alloy [24,25]. To reveal the effect of synergetic effects of Sn and Si, by keeping Zr-Va at either 1 NN or 2NN sites, we constructed and then calculated the binding energies of 54 Zr-Sn-Si-Va tetramers from the vast amounts of possible configurations (Table S4 to S5). To our surprise, most configurations exhibit significantly higher binding energies than that of Zr-Sn-Va trimers (Fig. 4a), showing maximum 0.33 eV and 0.41 eV for 1NN and 2NN Zr-Va constructions (Fig. S10), respectively, proving the synergetic effect of Sn and Si in catalyzing concentration fluctuations of Zr.

In addition to the increased chance for Zr to bind vacancy at 1 NN by forming low energy Zr-Sn-Va and Zr-Sn-Si-Va atom complexes, our calculations (Fig. 4b and S11) indicate that the migration energy barrier between Zr atom and 1NN vacancy is significantly reduced by forming Zr-Sn-Va, Zr-Si-Va and especially Zr-Sn-Si-Va atom complexes (from 0.97 to 0.62 eV atom$^{-1}$), implying a significantly increased diffusivity of Zr. It is likely that the diffusion of Zr was through the co-diffusion with Sn and/or Si atoms bounded in the atom complexes, via the "amoeba" mechanism as proposed by Embury *et al* [26]. Such a fast co-diffusion mechanism was also observed by Mao et al. by KMC simulation in a Ni-Al-Cr system [27]. This also explains why an addition of both Sn and Si has better inoculation effect than solely adding Sn, where Si has played an important role in enhancing the diffusion kinetics of Zr.

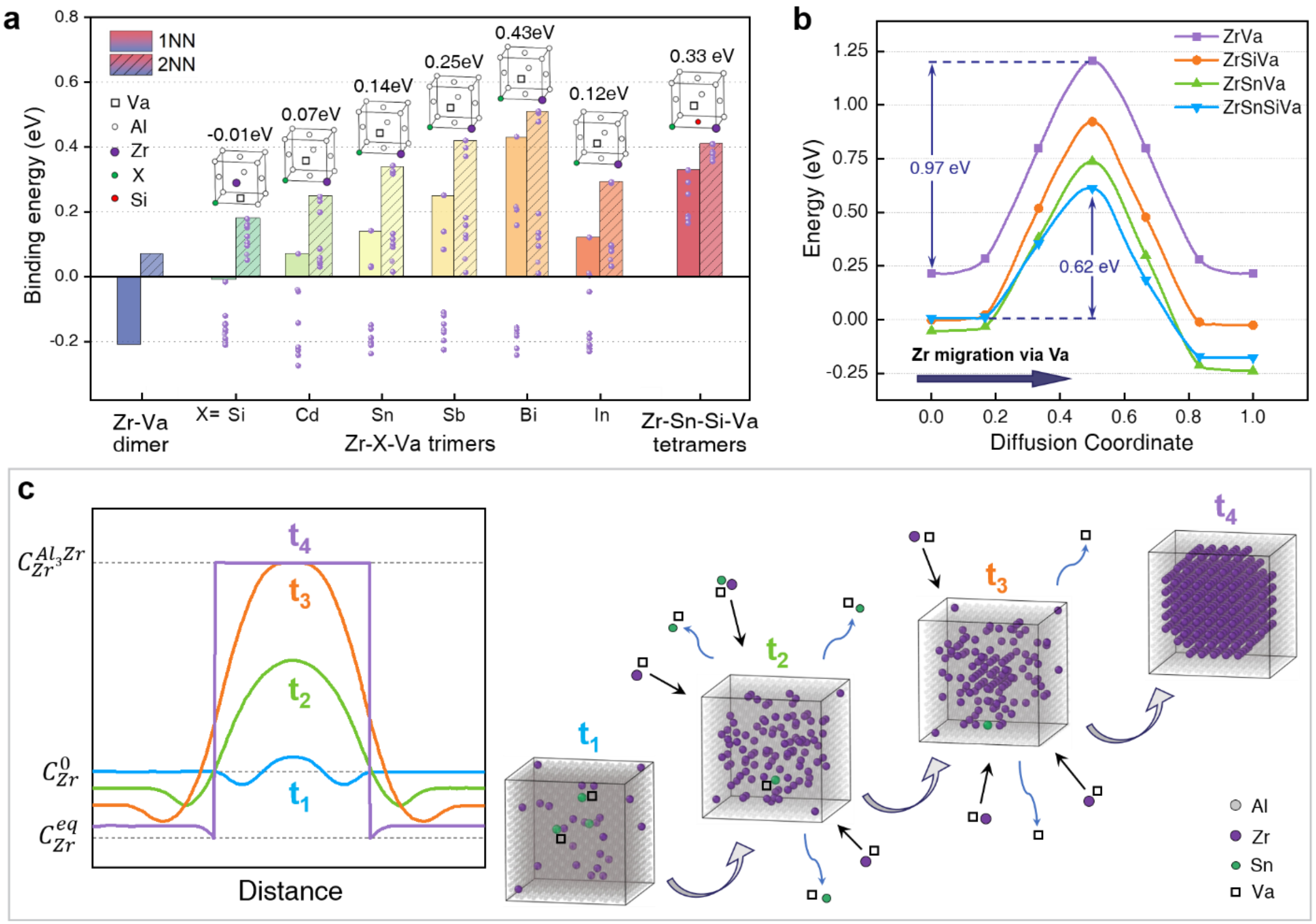

**Fig. 4. Influence of impurity inoculation on nucleation of $Al_3Zr$ through formation of Zr-rich atom clusters and diffusivity of Zr. a,** DFT calculated binding energies of Zr-Va dimer, Zr-X-Va trimers and Zr-Sn-Si-Va tetramers of different configurations (light purple points) where X represents impurity atom. The maximum values with Zr-Va at 1NN and 2NN are shown as columns with the corresponding schematic configurations of 1NN Zr-Va. **b,** Migration energy barrier of Zr atom calculated by CI-NEB when exchanging with 1NN Va in Zr-Va, Zr-Si-Va, Zr-Sn-Va and Zr-Sn-Si-Va complexes with the highest binding energies. **c,** Schematic drawing of the non-classical nucleation pathway of $Al_3Zr$ in ternary Al-Zr-Sn alloy, from local concentration fluctuation of Zr in forms Sn-Zr-Va atom complexes (t1) to Zr-rich clusters (t2), further growing into critical nucleus of $L1_2$-structured $Al_3Zr$ nuclei within cluster (t3) and finally to $Al_3Zr$ with sharp boundary to Al matrix.

Aided by the schematic drawings in Fig. 4c now we can draft the precipitation mechanism of $Al_3Zr$ precipitates in Al-Zr-Sn-Si alloy and describe it as follows. ***i)*** Spontaneous formation of low energy Zr-Sn-Si-Va atom complexes within the homogeneous solid solution; ***ii)*** Formation of local concentration fluctuation of Zr atoms around the (Zr, Sn, Si)-rich atom complexes; ***iii)*** Continuous increase of the amplitude in concentration fluctuation and CSRO of Zr in Zr-rich clusters, until forming $L1_2$-structured $Al_3Zr$ nuclei; ***iv)*** Growth of $Al_3Zr$ by consuming the surrounding Zr atoms and gradually rejecting the Sn and Si atoms. The pre-aging process has the influence of enhancing the formation of high number density of local concentration fluctuations of Zr. Since the Zr-rich clusters formed in the early stage have diffusive boundaries with surrounding Al matrix and lower concentration of Zr at periphery, the interfacial energy to the matrix, as the major nucleation barrier for $Al_3Zr$, can be substantially reduced. This precipitation kinetics is, to a large extent, analogous to the ***spinodal decomposition*** of unstable alloys. The major difference is that spinodal decomposition starts with continuous concentration fluctuation of small in amplitude and large in scale, while the present precipitation process starts with nanosized local Zr concentration fluctuations on low energy (Zr, Sn, Si)-rich complexes, which is discontinuous and of heterogeneous in nature. This is obviously a non-classical nucleation and growth pathway. The extremely dilute solid solution of normal binary Al-Zr alloys is known to be metastable in nature, which is resistant to any fluctuations. Obviously, the addition of Sn or both Sn and Si in Al-Zr alloys has changed the solid solution into an unstable state, where infinitesimal local fluctuation caused by formation of Zr-Sn-Va or Zr-Sn-Si-Va atom complexes can lower the free energy of the system. A further growth of the atom complexes into Zr-rich atom clusters with concentration gradient of Zr and diffuse interface with Al matrix is also supposed to be energetically favorable. It should be mentioned the expediting roles of early-stage atom clustering on the precipitation have long been noticed, and different mechanisms either through continuous transition from atom clusters to precipitates [28] or acting as nucleation sites for precipitates [29], have been discussed before. However, using impurity atoms to trigger the heterogeneous nucleation of precipitates through a spinodal decomposition like mechanism has never been reported. Here, in addition to providing catalytic sites for concentration fluctuation of Zr, the addition of impurity Sn and Si also significantly increases the diffusivity of Zr atoms in Al by reducing the migration energy barrier when exchanging position with 1NN vacancies in low energy Zr-Sn-Va and Zr-Sn-Si-Va atom complexes. This mechanism is especially important for alloys which use low diffusivity elements to form thermally stable nano precipitates to sustain a hight strength at elevated temperatures. These alloys generally suffer from sluggish precipitation kinetics.

Considering the low addition level of Sn, 0.02 at%, the inoculation effect realized in the alloy is huge. To explore similar highly potent inoculation impurity elements X for Al-Zr alloys, we generalize the following selection criteria of X, which ***i)*** has low solubility in Al solid solution in low temperature range (200-400 $^{o}$C) and therefore strong driving force to precipitate, ***ii)*** will not form Al-X intermetallic phase, ***iii)*** has higher diffusivity than Zr and ***iv)*** can form low energy Zr-X-Va or Zr-X-Si-Va atom complexes. Following above criteria, we have calculated the binding energies of Zr-X-Va trimers for most possible impurity elements in Al alloys. From the calculated results (Extended Data Table 4 and Fig S12), we could identify more impurity elements fulfilling above criteria, for example Cd, Sb, In and Bi (Fig. 4a), and tested Al-Zr-Sb-Si and Al-Zr-Bi-Si alloys. As shown in Extended Data Fig. 6, significantly enhanced precipitation hardening kinetics was achieved in both alloys, showing the validity of the above criteria.

In summary, we have introduced a simple but powerful strategy to inoculate the homogeneous solid-state precipitation of strengthening precipitates in metallic alloys by using trace level of impurity atoms to catalyze local concentration fluctuation, enhancing the diffusivity of solute elements by co-diffusion mechanism, and therefore realizing substantially faster precipitation kinetics. The precipitation is altered from the classical nucleation and growth pathway to a spinodal-decomposition-like non-classical nucleation pathway in a heterogeneous manner.

We hope the impurity atom inoculated precipitation of $Al_3Zr$ precipitates presented in this work could bring new insights into the nucleation and growth mechanisms of solid-state homogeneous precipitation of nano precipitates, while the proposed inoculation strategy can be extended to a broader range of metallic alloy systems, facilitating the exploration and development of more advanced precipitation-strengthened metallic materials.

# Method

## Material preparation

The Al-Zr based alloys studied in this work, including Al-Zr, Al-Zr-Sn, Al-Zr-Sn-Si, Al-Zr-Sb-Si and Al-Zr-Bi-Si, were prepared by the conventional gravity casting process and their detailed chemical compositions were measured by inductively coupled plasma-mass spectrometry (ICP-MS). **Extended Data Table 1** provides the measured chemical compositions. All experimental alloys were prepared by melting and mixing of 99.99% pure Al, pure Bi, Al-8Zr, Al-20Sn, Al-10Sb and Al-40Si master alloys (wt.%) in a crucible furnace, holding at 780 $^{o}$C for 40 min with regular stirring. After degassed by high-purity argon gas at 720 $^{o}$C, the liquid metal was poured into a copper mold to obtain ingots with dimensions of 25×70×100 $mm^3$.

Cubic samples (10×10×10 $mm^3$) cut from the ingots were aged directly from the as-cast state. No solution heat treatment was performed so as to avoid formation of coarse $Al_3Zr$ particles within dendrites. Various aging treatments were applied on the as-cast samples to investigate the precipitation behavior and the strengthening effect of $Al_3Zr$ nanoparticles, including:

(1) Multi-steps isochronal aging in air furnace over the temperature range of 150-550 $^{o}$C with 50 $^{o}$C increments and a 1 h holding time at each temperature;

(2) Isothermal aging treatments at 400 $^{o}$C for times up to 500 h;

(3) Two-step aging treatments, designed to promote the nucleation kinetics, were carried out on a sequence of pre-aging either at 200 °C for 24 h or at 300 °C for 8 h (Step-1) followed by 400 $^{o}$C isothermal aging for various times up to 500 h (Step-2).

## Electrical conductivity and mechanical testing

The precipitation behavior during aging was monitored by measurements of Vickers microhardness and electrical conductivity (EC) at room temperature upon different aging states. The hardness values were measured by utilizing an Innovatest machine with a load of 200 g and 15 s dwell time on sample surface polished to 1 µm. At least ten indentations across different grains were measured for each sample. SigmaTest 2.069 (Foerster, Germany) was used to measure EC of samples by using eddy currents at five frequencies (60, 120, 240, 480 and 960 kHz). Dog-bone shaped specimens with gauge dimensions of 30×6×2 $mm^3$ were machined from the as-cast ingots (as-per ASTM standard), and then are subjected to isochronal aging to the peak-aged states based on the hardness measurement results. Tensile tests were carried out at room temperature using a strain rate of $1\times10^{-3}$ $s^{-1}$. At least three specimens were tested for each alloy to ensure the reproducibility.

## Microstructure observations by electron microscopies

Samples for scanning electron microscopy (SEM) analysis were first prepared by standard grounding and polishing methods, and then electropolished in electrolyte of 1/3 nitric acid in methanol at -30 °C and 25V for 45 s. Micrographs were taken at BSE mode using a low voltage of 5 kV and a working distance of 6 mm. The contrast of the matrix grain was finely tuned by tilting the stage to meet the two-beam condition (i.e. the so-called channeling contrast) to signify the contrast from the nanoprecipitates.

For transmission electron microscopy (TEM) observations, samples were firstly sliced into 0.5 mm thick foils via wire electrical discharge machining (EDM) and ground to ~100 µm thickness. Subsequently, foils were punched into 3 mm diameter discs and were subjected to twin-jet electropolishing using Struers Tenupol-5 apparatus in an electrolyte of 1/3 nitric acid in methanol at -25 °C and 20V. Prior to observations, the thinned discs were plasma cleaned by Fishione 1051 system. The conventional selected area electron diffraction (SAED) and dark-field imaging were operated at 200 kV on JEOL 2010/2100 TEMs.

A Thermo Fisher Titan Themis-Z aberration-corrected microscope at 300 kV was used to take high resolution high-angle annular dark-field (HAADF-) scanning transmission electron microscopy (STEM). A 15 mrd convergence semi-angle and a beam current of 76 pA were used. The STEM micrographs were collected in the semi-angle of ~64 mrad. This imaging mode is particularly sensitive to the atomic number Z under the beam and thus is very conducive to identifying small atomic clusters in the present Al-Zr system [30]. In addition, the atomic resolution EDS mapping was performed at 200 kV using Super-X energy dispersive x-ray spectroscopy (Super-X EDS system). We conducted quantitative image analysis using Image J for the measurement of the size and density of the $Al_3Zr$ nanoparticles.

**APT experiments and data analysis**

Nanotips for three-dimensional (3D) atom-probe tomography (APT) investigations were prepared by cutting ~0.4×0.4×10 $mm^3$ blanks from the aged samples, followed by a two-step electropolishing technique [31]. APT experiments were conducted in laser mode utilizing a picosecond laser-pulsed Local Electrode Atom Probe (LEAP 5000XS) from CAMECA Inc. The temperature was set at 30 K in ultrahigh vacuum below $2.0\times10^{-9}$ Pa. The laser kept a pulse repetition rate of 500 kHz and the energy was calibrated for each individual tip to yield an equivalent pulse fraction in voltage mode of 25%, which corresponded to laser energy between 60 and 150 pJ. The detection rate was set to have 0.5% of the applied laser pulses resulting in an evaporation event.

The acquired datasets were reconstructed and analyzed using APSUITE™ 6 software. Herein, similar to ref. [25], Si evaporates mainly as $Si^{2+}$, whose peaks in the mass spectrum can be clearly identified at the decay tail of the $Al^{2+}$ peak, thus ensuring the accuracy of selected Si ions. Besides, Zr evaporates exclusively as $Zr^{2+}$, while a very small amount of $Zr^{3+}$ were not selected due to its overlap with the tail of $Al^{+}$ peak. The size and number density of precipitates and solute clusters were quantified based on cluster identification from iso-position method [32,33]. The proximity histogram methodology [34] based on iso-surfaces was utilized to study the compositional variations within the nanoprecipitates and the matrix, where background corrections were performed to improve the accuracy of the compositional measurements. To detect atom complex/clusters and short range ordering of solute atoms formed in the pre-aging treatments, the partial radial distribution function (PRDF) technique [25,35,36] was employed. A PRDF at a radial distance $r$ is defined as the average concentration distribution of $P$ atoms around a given solute species $V$, $< C_P^V(r) >$, normalized to the overall concentration of $P$ atoms $c_P^0$, in the sampled volume:

$$\text{PRDF} = \frac{< C_P^V(r) >}{c_P^0} = \frac{1}{c_P^0} \sum_{k=1}^{N_V} \frac{N_P^k(r)}{N_{tot}^k(r)}$$

where $N_P^k(r)$ is the number of $P$ atoms in a radial shell of unit thickness around the $k^{th}$ $V$ atom that is at the center of a shell with radius $r$, $N_{tot}^k(r)$ is the total number of atoms in this shell, $N_V$ is the number of $V$ atoms in the analyzed volume. PRDF values of unity indicate perfect random distribution, and PRDF values larger than unity describe clustering of the species $P$ and $V$.

## Calculation of the thermodynamics of precipitation

The nucleation thermodynamics of $Al_3Zr$ precipitates in binary Al-Zr alloys is evaluated by the classic nucleation theory. The solute concentration of Zr before aging treatment is assumed to be the same as the nominal concentration of experimental alloy, $C_0$= 0.08 at.%, The equilibrium solubility of Zr Al solid solution is expressed as$C_{Zr}^{eq} = \exp[(-0.620 + 155 \times 10^{-6}T)\ eV/k_bT]$ [10], where $k_b$ is the Boltzmann constant (with unit of eV/K) and T is temperature.  The diffusivity of Zr in is solid solution is given by $D_{Zr} = 728 \times 10^{-4} \exp(-2.51eV/k_bT)$. The critical nucleation energy barrier of $Al_3Zr$ is calculated by

$$\Delta G^* = 16\pi\gamma^3/3(\Delta G_v + \Delta G_\varepsilon)^2$$

Where $\gamma$ is the interfacial energy between $Al_3Zr$ and Al. $\Delta G_v$ and $\Delta G_\varepsilon$ are the volumetric Gibbs free energy change and the elastic strain energy for formation of $Al_3Zr$ phase, respectively, expressed as [4]

$$\Delta G_v = \frac{C_{Zr}^{\beta} R_g T}{V_\beta} \ln\left(\frac{C_0}{C_{Zr}^{eq}}\right)$$

$$\Delta G_\varepsilon = 8\mu \frac{(a_{Al} - a_\beta)^2}{(a_{Al} + a_\beta)^2}\left(\frac{1 + \mathcal{V}}{1 - \mathcal{V}}\right)$$

Where $V_\beta$ is the molar volume of $Al_3Zr$ , $C_{Zr}^{\beta}$is the Zr concentration in $Al_3Zr$ , $C_{Zr}^{\beta}$ =0.25, $R_g$ is gas constant, $\mu$ the shear modulus of Al ( $\mu = 25.0GPa$), $a_{Al} = 0.405\ nm$ and $a_\beta = 0.4109\ nm$ are lattice constant of Al and $Al_3Zr$, respectively. $\mathcal{V}$ is Poisson number ($\mathcal{V} = 0.33$).

The incubation time τ for steady state nucleation is calculated by $\tau = \frac{1}{2Z^2\beta^*}$ [20] or $\tau = \frac{2}{3\pi Z^2\beta^*}$ [20]

where

$$\beta^* = \frac{16\pi\gamma^2 D_{Zr} C_0}{a^4 {\Delta G_v}^2} \text{ and } Z = \frac{V_a {\Delta G_v}^2}{8\pi(k_b T\gamma^3)^{\frac{1}{2}}}$$

In the equations $a$ is the lattice distance of Al matrix, $a = 0.287nm$; Z is *Zeldovich* factor; $V_a$ is the molar volume of Al matrix.

The growth rate of Zr-rich atom clusters during constant temperature heating is estimated using following equation [4],

$$r_t = (2D_{Zr}t\,\frac{C_{Zr} - C_{Zr}^{eq}}{C_{Zr}^{p} - C_{Zr}^{eq}})^{\frac{1}{2}}$$

where $r_t$ is the radius of atom cluster as a function of heating time $t$, $C_{Zr}^{p}$ is the average concentration of Zr in atom cluster.

**First principles calculation**

First-principles density functional theory (DFT) calculations were carried out with the Vienna Ab initio Simulation Package (VASP) [37,38], using the projector augmented wave (PAW) potentials [39,40]. The exchange-correlation energy was described by the Perdew-Burke-Ernzerhof (PBE) parameterization of generalized gradient approximations (GGA) [41]. Brillouin zone integration was made with the first order Methfessel-Paxton smearing [42] with a width of 0.2 eV. The structures were fully relaxed at a plane-wave cutoff energy of 520 eV until the total free energy and atomic forces were converged to $1\times10^{-6}$ eV and $1\times10^{-2}$ eV/Å, respectively. The calculations were performed with a 4×4×4 conventional FCC supercell (total atomic sites N = 256). Gamma-centered k-point meshes were used to achieve approximately 16000 k-points per reciprocal atom. Calculations with different cell size and k-point density indicate that the binding energies were converged to ± 0.03 eV. For calculation of vacancy migration barriers, the climbing image nudged elastic band (CI-NEB) method was used [43,44]. OVITO [45] was used to make schematic drawing of solute clusters in Fig. 4c.

The binding energies of vacancy containing atom complexes, including dimers, trimers and tetramers, were defined as the difference between the sum of chemical potentials for all atoms in solid solution and the calculated total energy of the simulated box (positive binding energy indicates stability):

$$E_{bind} = -\left(E_{tot} - \sum_{i=1}^{N} \mu_M^{i,SS}\right), M = Al, Zr, Va, X$$

where $E_{tot}$ is the total energy of the simulated box, $N$ is the number of atomic sites in the box, and $\mu_M^{i,SS}$ is the chemical potential of the atom at lattice site $i$, for a certain element $M$ (Va stands for vacancy) in solid solution (SS):

$$\mu_{Al}^{i,SS} = E_{tot}(Al_N)/N$$

$$\mu_Y^{i,SS} = E_{tot}(Al_{N-1}Y_1) - (N-1)\mu_{Al}, Y = Zr, Va, X$$

Configurations of Zr-X-Va trimers were built on based on Zr-Va dimer at 1NN or 2NN, with either Zr or Va as the center atom and the X atom positioned at the 1NN or 2NN sites of the center atom were exhaustedly enumerated with the help of the Atomic Simulation Environment (ASE) [46] and the binding energy, $E_{bind}$, of each configuration was calculated with DFT (Extended Data Fig. 5 and Table 6). All the 5 migration energy barriers for exchange between Zr and 1NN Va were then calculated in the case of Zr-Sn-Va trimers.

For binding energy of Zr-Sn-Si-Va tetramers, the same enumeration was performed, but the number of configurations was too high that it was computationally prohibitive to perform DFT calculations on all of them. Instead, a rough estimation of the binding energy was performed on each of the structure based on the binding energy for dimers (as shown in Table S4 and S5):

$$E_{bind} \approx \sum_{i,j} E_{bind}^{k}(i-j)$$

where $i, j$ = Zr, Sn, Si, Va is one of the components of tetramer and $i, j$ are in $k^{th}$ NN distance with each other, and interactions beyond 4NN were set to zero. DFT calculations were performed on the tetramer structures with Zr-Va at 1NN and binding energy larger than 0.14 eV by the above estimation, as shown in Table S4, and 2NN Zr-Va structures with binding energy larger than 0.4 eV by estimation, as shown in Table S5. Since the configuration with highest binding energy should dominate diffusion, Zr-vacancy migration barriers for 14 tetramers with DFT-calculated binding energy more than 0.10 eV were then calculated.

The impurity substitution energy (Table S3) is calculated as

$$E_{sub}^{Al_3Zr} = E_{tot} - \sum_{i=1}^{N} \mu_{M}^{i,Al_3Zr}, M = Al, Zr, Va, X$$

where $\mu_{M}^{i,Al_3Zr}$ is the chemical potential of the species in the presence of $Al_3Zr$ precipitates:

$$\mu_{Y}^{i,Al_3Zr} = \mu_{Y}^{i,SS}, Y = Al, Va, X$$

$$\mu_{Zr}^{i,Al_3Zr} = \frac{4}{N} E_{tot}(Al_3Zr) - 3\mu_{Al}$$

**Acknowledgements:** The Research Council of Norway (RCN) is acknowledged for funding SFI Physmet (project number 309584) and the NTNU atom probe facility through the Norwegian Laboratory for Mineral and Materials Characterization (MiMaC) project number 269842. The DFT calculations were performed on the Sigma2 high performance computer clusters (Grants No. nn9347k).

**Author contributions:** Y.L. developed the concept, designed the experiments and supervised the project. S.P. prepared all samples, performed mechanical tests and preliminary conventional TEM. H.S. and C.H. performed the APT. F.Q. and S.P. performed all atomic-resolution transmission electron microscopy; C.L. and D.Z. performed the DFT calculation; H.S., S.P. and C.L analyzed the results. Y.L., S.P. and C.L. wrote the paper.

**Competing interests:** Authors declare no competing interests.

**Data availability:** All data to evaluate the conclusions are present in the manuscript, the Extended Data items and the Supplementary Information. Raw data are available from the corresponding authors on reasonable request.

**Additional information:**

**Supplementary information:** Supplementary Information is available for this paper.

**Correspondence and requests for materials** should be addressed to Yanjun Li.

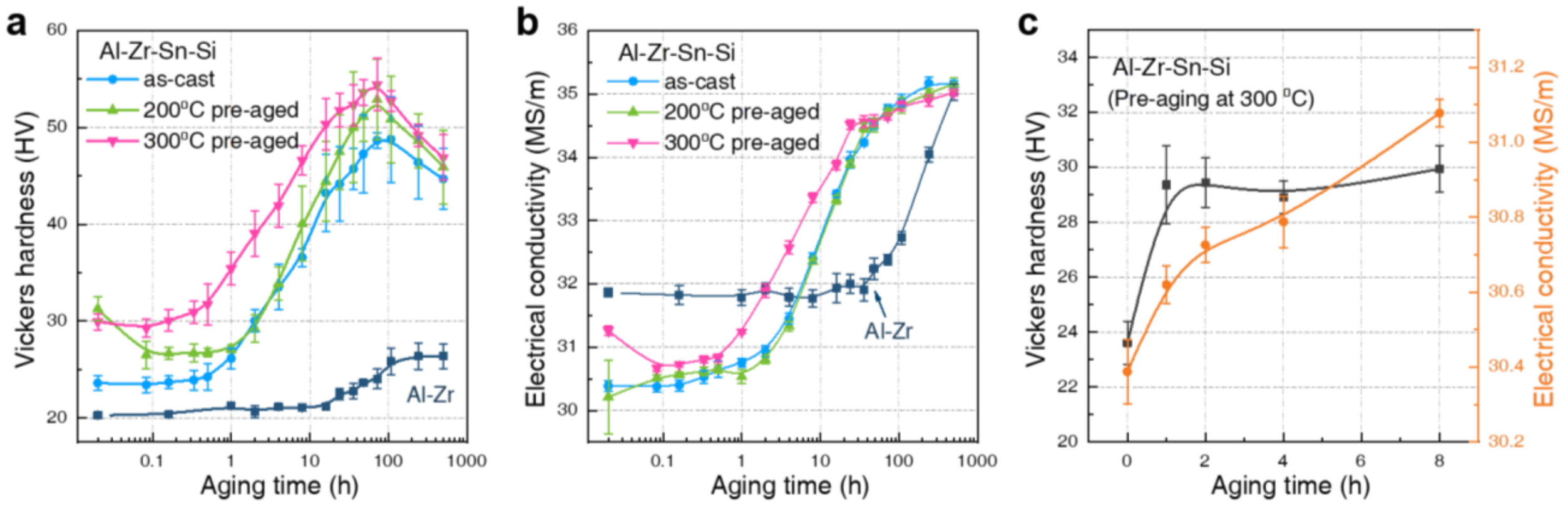

**Extended Data Fig. 1 | Evolution of microhardness and electric conductivity (EC) of Al-Zr-Sn-Si alloy during two-step aging. a-b,** Measured values at room temperature after 400 °C aging for different times, starting either from as-cast state, or after pre-aging at 200 °C for 24 h or at 300 °C for 8h. The hardness and EC evolution of Al–Zr is also included for comparison. **c,** Variation in hardness and EC of Al-Zr-Sn-Si during pre-aging at 300 °C until 8h. Compared with the as-cast sample, both the hardness and EC values of the pre-aged samples are remarkably higher, which is attributed to the formation of nano-sized Sn-rich clusters. During subsequent 400 °C aging, the pre-aged samples also exhibited faster precipitation kinetics and obtained higher peak-state hardness values. The remarkable increase of hardness during 300 °C pre-aging shows the strong inoculation effect of nano-sized Zr-rich atom clusters.

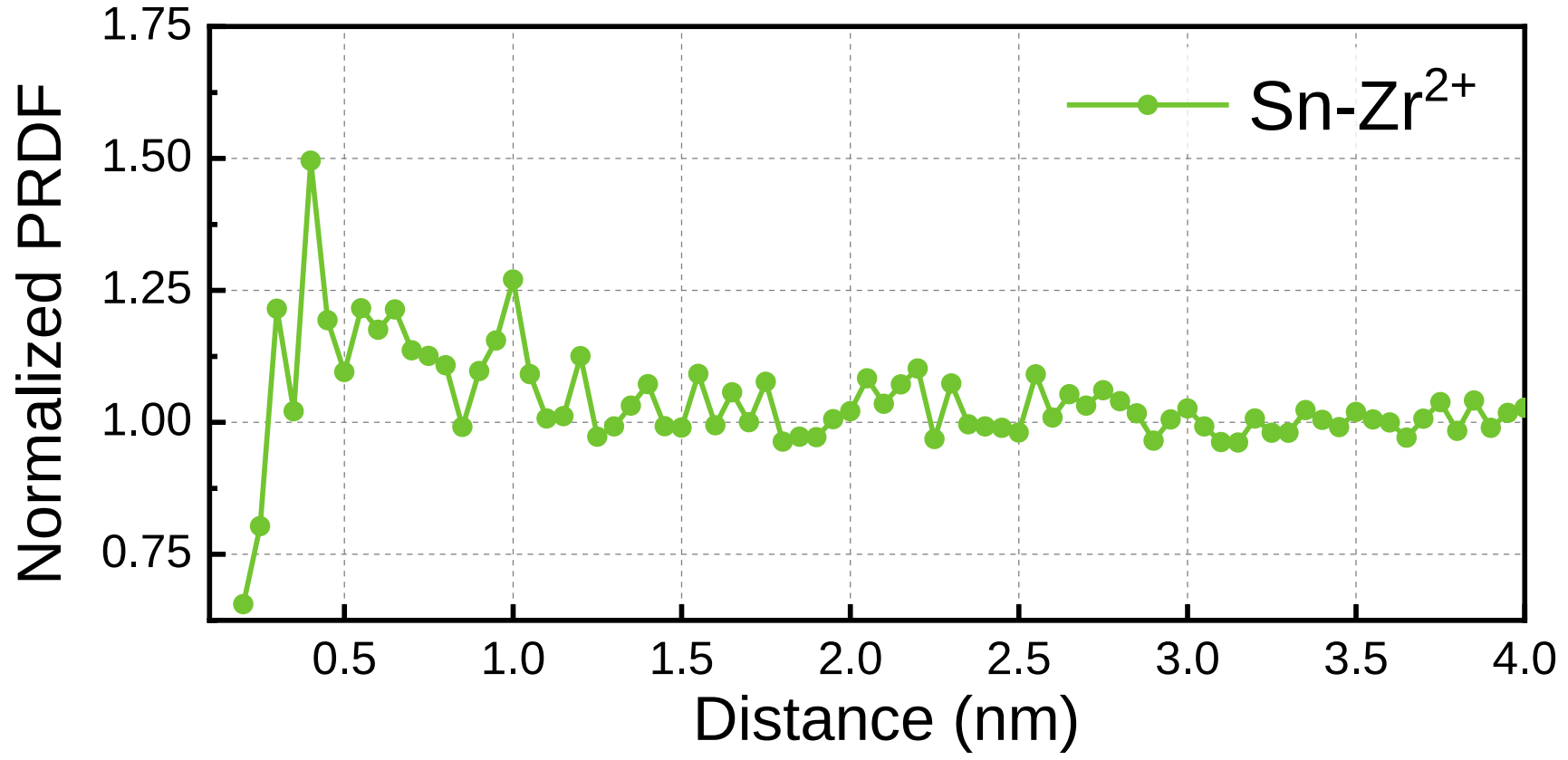

**Extended Data Fig. 2 | Normalized partial radial distribution function (PRDF) of $Zr^{2+}$-Sn after excluding all Sn-rich particles in the reconstructed tip in Fig. 2a.** A weak Sn-Zr clustering can be detected within the range of distance <1 nm, showing the short-range ordering between Sn and Zr atoms.

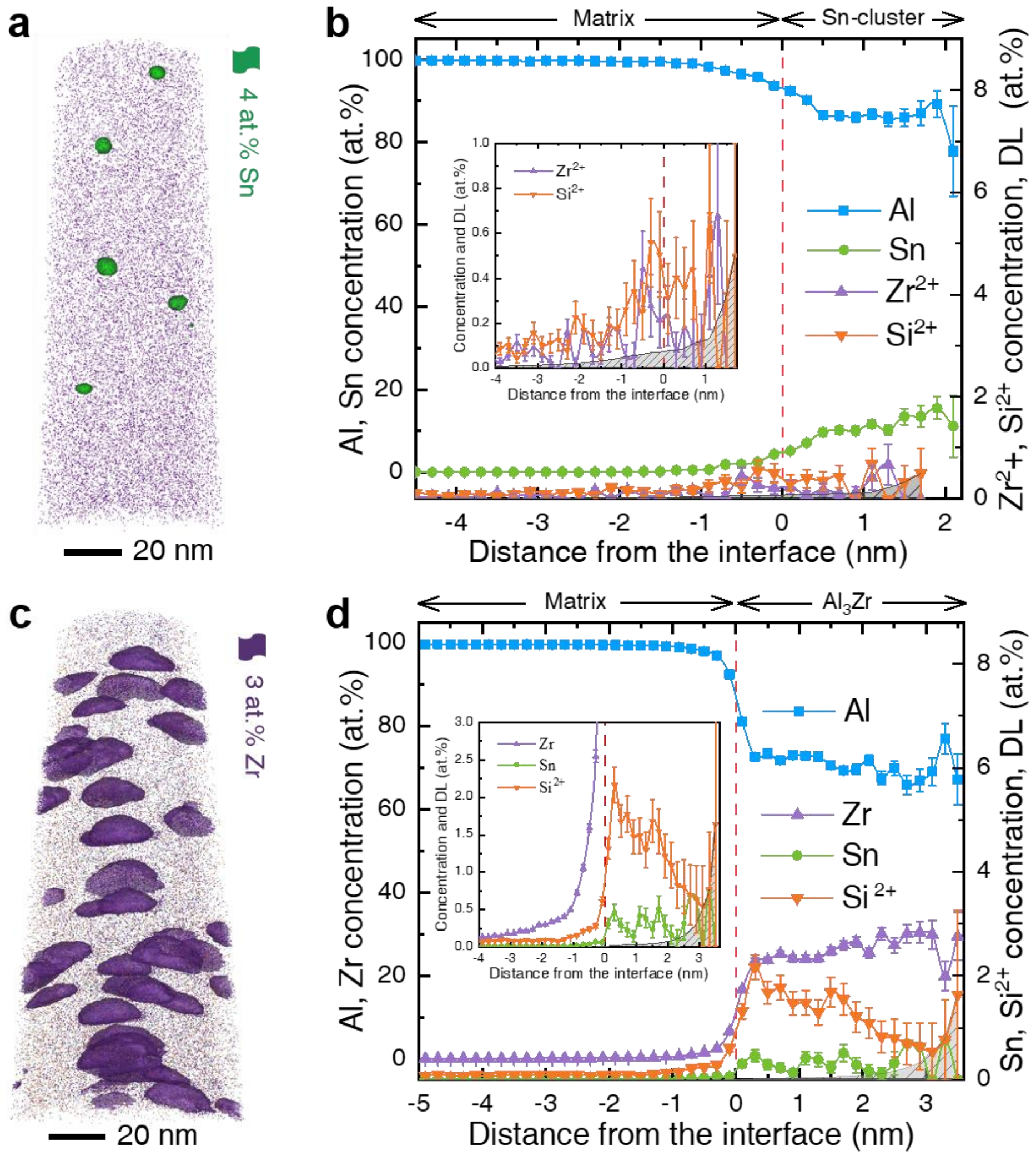

**Extended Data Fig. 3 | Atom probe analysis of Al-Zr-Sn-Si alloy after 200 ºC/24 h preaging and after peak aging (200 ºC/24 h + 400 ºC/72 h). a,** Reconstruction of pre-aged sample to show the Sn-rich particles highlighted by an iso-surface encompassing regions containing more than 4 at% Sn. **b,** Proximity histogram of Sn-rich atom clusters, showing both Zr and Si atoms are partitioned in the clusters. **c,** Reconstruction of peak-aged sample with $Al_3Zr$ delineated in 3 at% Zr iso-surfaces. **d,** Proximity histogram of the $Al_3Zr$ precipitates, showing significant amount of both Si and Sn atoms are retained in the precipitates, which may be attributed to the nucleation of $Al_3Zr$ precipitates from (Sn, Zr, Si)-rich atom clusters. Note the matrix/precipitate interface (vertical red dot-dash line) is defined as the inflection point of the Al concentration profile. The gray shaded areas represent the detection limit (DL) defined as one atom per proxigram bin (in at.%).

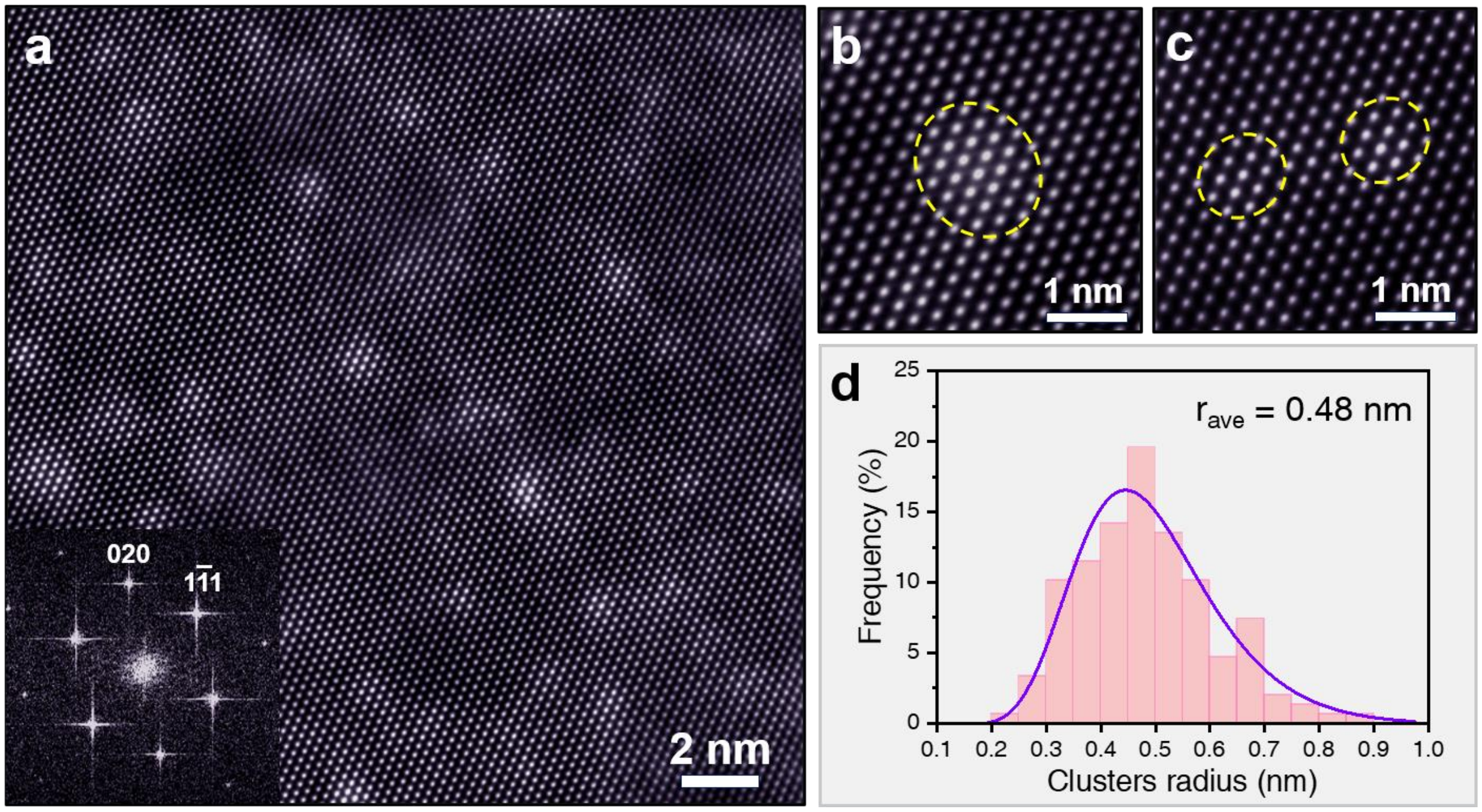

**Extended Data Fig. 4 | Atomic-resolution HAADF-STEM images of the 300 ºC pre-aged Al-Zr-Sn-Si alloy viewing along <011>$_{Al}$ projection.** In **a,** a large number of nano-sized features with higher mass contrast can be observed. However, the fast Fourier transformation (FFT) pattern reveals no extra diffraction spots other than Al matrix showing the features are atom clusters enriched with heavy elements Zr and/or Sn. **b-c,** enlarged solute clusters, which show a short-range ordered structure instead of the long range $L1_2$-structure ordering of $Al_3Zr$ phase. **d,** Statistical size distribution of the observed clusters measured with HAADF-STEM images, showing an average radius of ~0.48 nm.

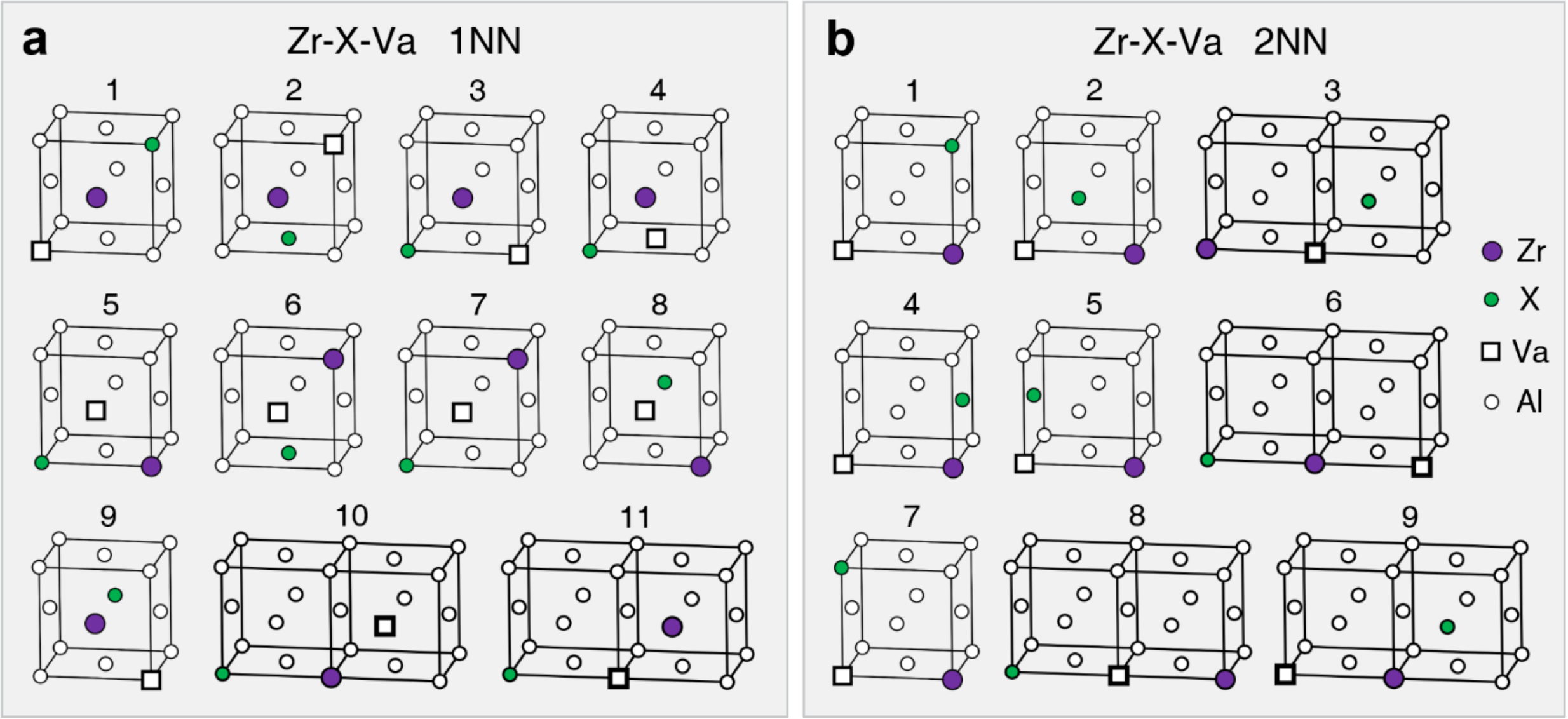

**Extended Data Fig. 5 | Atomic configurations for Zr-X-Va trimers within the Al supercells with X representing impurity atom and Va representing vacancy.** Note Zr and Va are positioned at 1NN in **a** or 2NN in **b**. All possible configurations, including 11 for 1NN and 9 for 2NN are exhaustively enumerated, with X atom at 1NN or 2NN position to either Zr or Va. Only the corners of respective 256-atoms FCC Al supercell (4×4×4) are displayed.

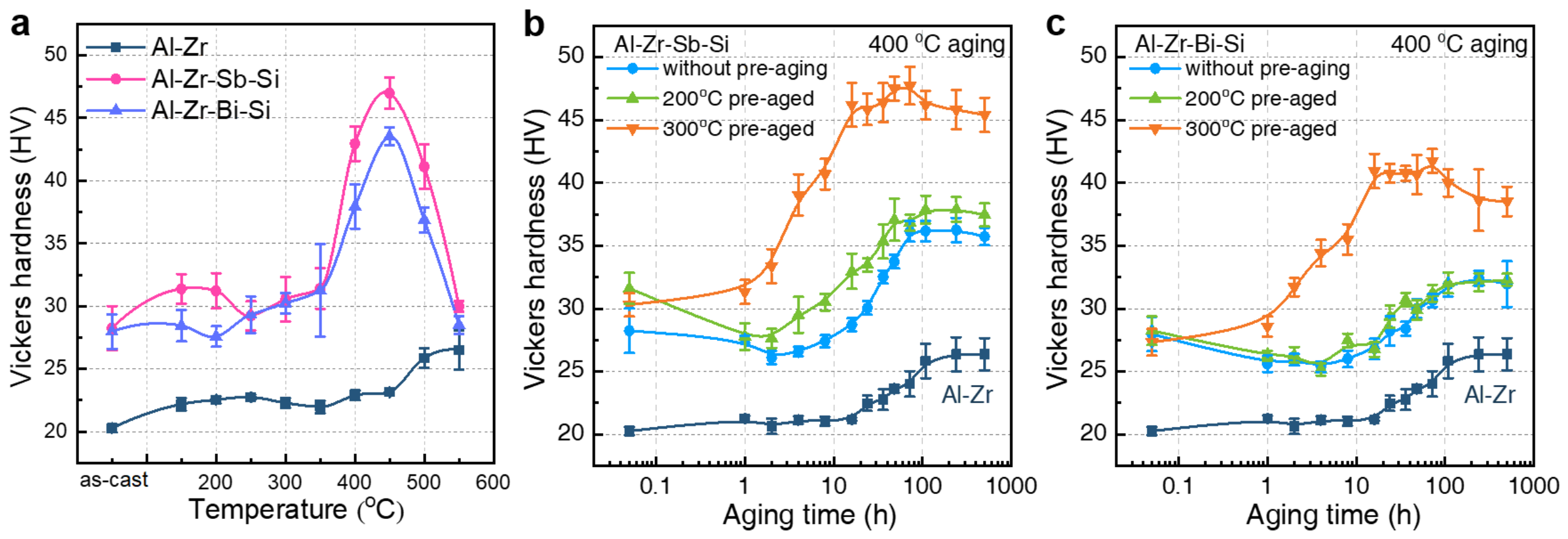

**Extended Data Fig. 6 | Evolution of microhardness of Al-Zr-Sb-Si and Al-Zr-Bi-Si alloys during various aging regimes. a,** Microhardness evolution during multi-steps isochronal aging from 25 to 550 °C. The precipitation and hardening of $Al_3Zr$ can be effectively promoted via addition of Sb/Bi and Si impurity elements. **b-c,** Hardness variation during isothermal aging of Al-Zr-Sb-Si and Al-Zr-Bi-Si samples, with and without 200 °C/24 h or 300 °C/8h pre-aging, respectively. It proves that a suitable pre-aging treatment to promote clusters inoculation is essentially required for achieving "best-case" hardening potential.

**Extended Data Table 1 | Chemical compositions (at. %) of the as-cast studied alloys, as measured by Inductively Coupled Plasma Mass Spectroscopy (ICP-MS).**

| Alloys | Al | Zr | Si | Sn | Sb | Bi |
|---|---|---|---|---|---|---|
| Al-Zr | Bal | 0.080 | - | - | - | - |
| Al-Zr-Sn-Si | Bal | 0.078 | 0.110 | 0.015 | - | - |
| Al-Zr-Sn | Bal | 0.080 | - | 0.017 | - | - |
| Al-Zr-Sb-Si | Bal | 0.082 | 0.090 | - | 0.020 | - |
| Al-Zr-Bi-Si | Bal | 0.079 | 0.140 | - | - | 0.015 |

**Extended Data Table 2 | Tensile yield strength (YS), ultimate strength (UTS) and elongation (EL) of studied alloys at peak-aged states during isochronal aging.**

| Alloys | State | YS (MPa) | UTS (MPa) | EL (%) |
|---|---|---|---|---|
| Al-Zr | Isochronal aged to 500 ºC | 33.4 ± 1.6 | 64.4 ± 0.9 | 25.4 ± 3.3 |
| Al-Zr-Si-Sn | Isochronal aged to 450 ºC | 72.7 ± 1 | 116 ± 0.5 | 17.2 ± 0.4 |
| Al-Zr-Sn | Isochronal aged to 450 ºC | 64.1 ± 0.8 | 103.1 ± 2.7 | 14.5 ± 5.2 |
| Al-Zr-Si-Sb | Isochronal aged to 450 ºC | 57.8 ± 2.1 | 103.7 ± 4.6 | 22.4 ± 8.3 |
| Al-Zr-Si-Bi | Isochronal aged to 450 ºC | 48.9 ± 4.5 | 93.1 ± 2.6 | 24.2 ± 6.6 |

**Extended Data Table 3 | Calculated solute-vacancy (solute-Va) and solute-solute binding energies in Al from the first to the fourth NN distances.** Here, various solute atoms, including Zr, Sn and Si existing in the studied Al-Zr-Sn-Si alloy, were taken into consideration. Positive binding energies are attractive and therefore energetically favorable, and vice-versa.

| Different dimers | Binding energies (eV) | | | |
|---|---|---|---|---|
| | 1NN | 2NN | 3NN | 4NN |
| Zr-Va | -0.18 | 0.09 | -0.04 | 0.05 |
| Sn-Va | 0.25 | -0.02 | 0.01 | 0.01 |
| Si-Va | 0.06 | 0.00 | 0.01 | 0.00 |
| Zr-Sn | 0.08 | 0.05 | 0.04 | 0.00 |
| Zr-Si | 0.08 | 0.02 | 0.01 | -0.02 |
| Sn-Si | -0.01 | -0.01 | 0.00 | 0.01 |

**Extended Data Table 4 | Calculated binding energies (eV) in 1NN and 2NN for Zr-X-Va trimers with different configurations.** Herein, X refers to several different impurity atoms (Si, Sn, Sb, Bi, Cd, In) which are expected to favor the diffusion of Zr in Al matrix. The corresponding schematic figures of all different configurations are displayed in Extended Data Fig. 5.

| Configurations | Zr-Si-Va | | Zr-Sn-Va | | Zr-Sb-Va | | Zr-Bi-Va | | Zr-Cd-Va | | Zr-In-Va | |
|---|---|---|---|---|---|---|---|---|---|---|---|---|
| | 1NN | 2NN | 1NN | 2NN | 1NN | 2NN | 1NN | 2NN | 1NN | 2NN | 1NN | 2NN |
| 1 | -0.14 | 0.10 | -0.15 | 0.10 | -0.11 | 0.13 | -0.16 | 0.20 | -0.23 | 0.09 | -0.17 | 0.08 |
| 2 | -0.15 | 0.18 | -0.16 | 0.34 | -0.12 | 0.42 | -0.17 | 0.48 | -0.24 | 0.20 | -0.19 | 0.29 |
| 3 | -0.12 | 0.12 | -0.19 | 0.32 | -0.15 | 0.37 | -0.22 | 0.50 | -0.27 | 0.23 | -0.22 | 0.29 |
| 4 | -0.01 | 0.16 | 0.03 | 0.13 | 0.14 | 0.18 | 0.16 | 0.12 | -0.14 | 0.04 | -0.05 | 0.09 |
| 5 | -0.20 | 0.16 | 0.14 | 0.33 | 0.25 | 0.40 | 0.43 | 0.51 | 0.07 | 0.25 | -0.21 | 0.29 |
| 6 | -0.15 | 0.11 | 0.03 | 0.09 | 0.08 | 0.12 | 0.21 | 0.14 | -0.05 | 0.06 | -0.21 | 0.08 |
| 7 | -0.18 | 0.05 | 0.03 | 0.02 | 0.08 | 0.01 | 0.22 | 0.01 | -0.04 | 0.05 | 0.12 | 0.10 |
| 8 | -0.19 | 0.07 | -0.24 | 0.05 | -0.22 | 0.05 | -0.24 | 0.04 | -0.23 | 0.05 | 0.00 | 0.03 |
| 9 | -0.16 | 0.15 | -0.21 | 0.12 | -0.17 | 0.16 | -0.16 | 0.10 | -0.22 | 0.03 | 0.01 | 0.05 |
| 10 | -0.17 | - | -0.20 | - | -0.16 | - | -0.16 | - | -0.23 | - | -0.23 | - |
| 11 | -0.21 | - | -0.21 | - | -0.20 | - | -0.18 | - | -0.22 | - | -0.21 | - |

Supplementary Materials for

# Inoculating solid-state homogeneous precipitation by impurity atoms through a spinodal decomposition like pathway

Shiwei Pan [1,3], Chunan Li [1,3], Hanne-Sofie Søreide [1,3], Dongdong Zhao [1], Constantinos Hatzoglou [1], Feng Qian [1] ✉, Long-Qing Chen[2], Yanjun Li [1] ✉

[1] Department of Materials Science and Engineering, NTNU, Trondheim, Norway

[2] Department of Materials Science, Pennsylvania State University, Pennsylvania, USA

[3] These authors contributed equally.

✉ e-mail: qianfeng1101@163.com, yanjun.li@ntnu.no

**This file includes:**

Figs. S1 to S12

Tables S1 to S5

Movies S1 to S2

References

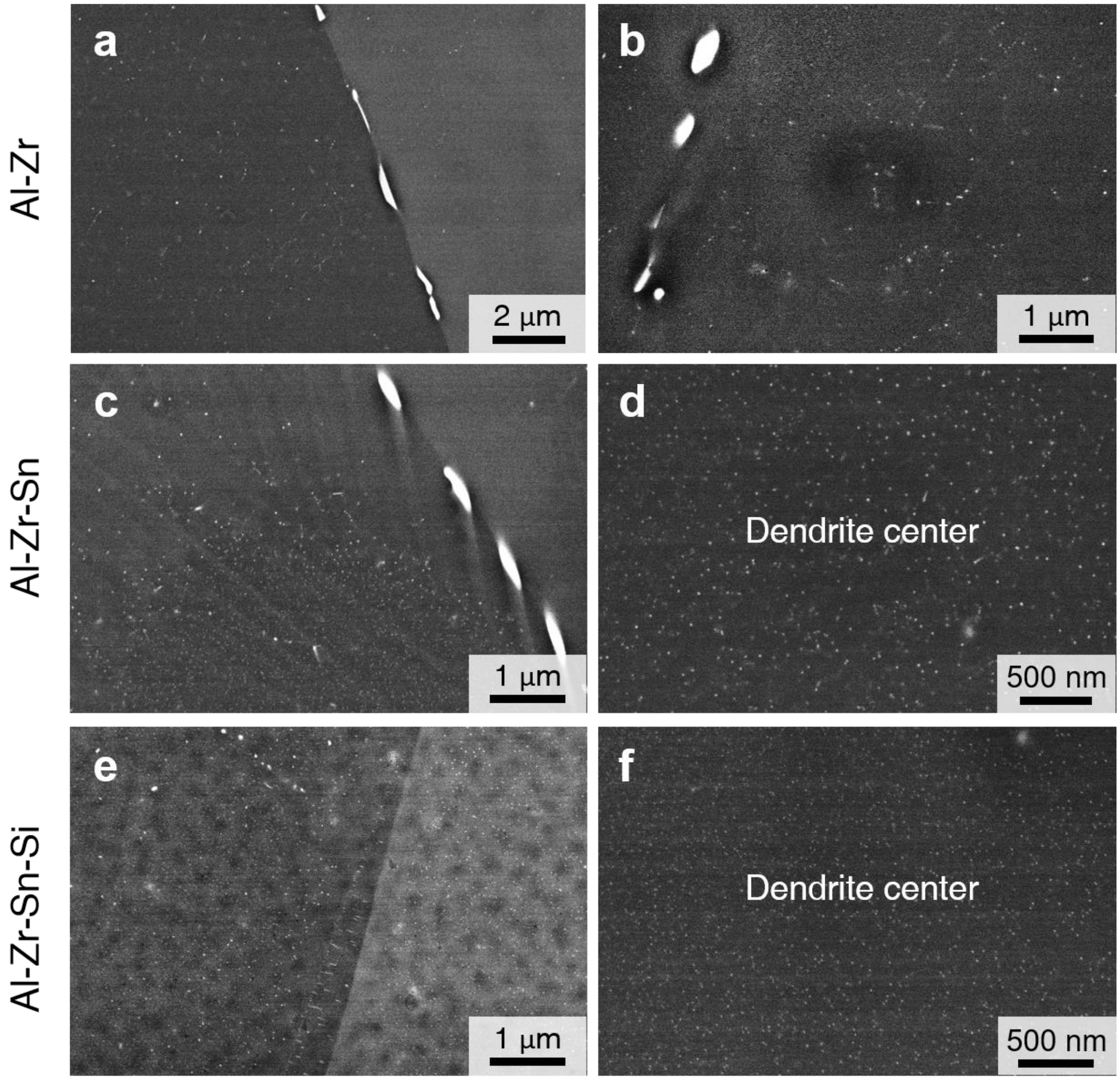

**Supplementary Fig. S1 │ High-resolution SEM-BSE micrographs of different alloys heat treated to peak-aged state by isochronal heating. a-b,** Binary Al-Zr alloy as-heated to 500 $^{o}$C. **c-d,** Al-Zr-Sn and **e-f,** Al-Zr-Sn-Si alloys as-heated to 450 $^{o}$C. All the images were taken at the two-beam condition and the bright dots are $Al_3Zr$ precipitates showing a higher Z-contrast than the Al matrix. The distribution of sparse $Al_3Zr$ in Al-Zr binary alloy, indicating a weaker precipitation kinetics and some local heterogeneous nucleation along dislocations. In contrast, Al-Zr-Sn and Al-Zr-Sn-Si alloy show much denser and finer $Al_3Zr$ precipitates homogeneously distributing across dendrite arms.

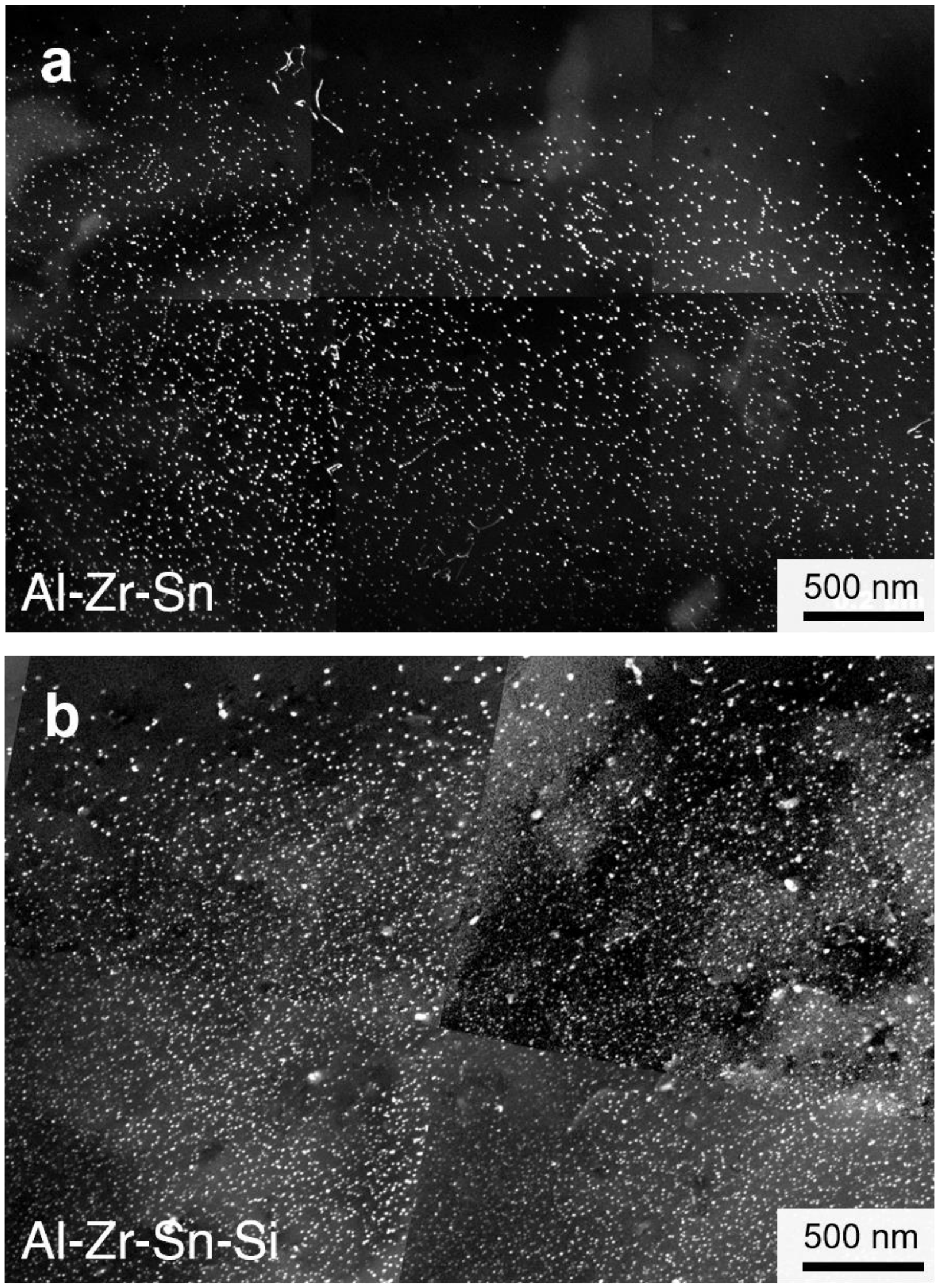

**Supplementary Fig. S2 │ The montaged TEM dark-field (DF) images of the peak-aged (a) Al-Zr-Sn and (b) Al-Zr-Sn-Si alloys.**

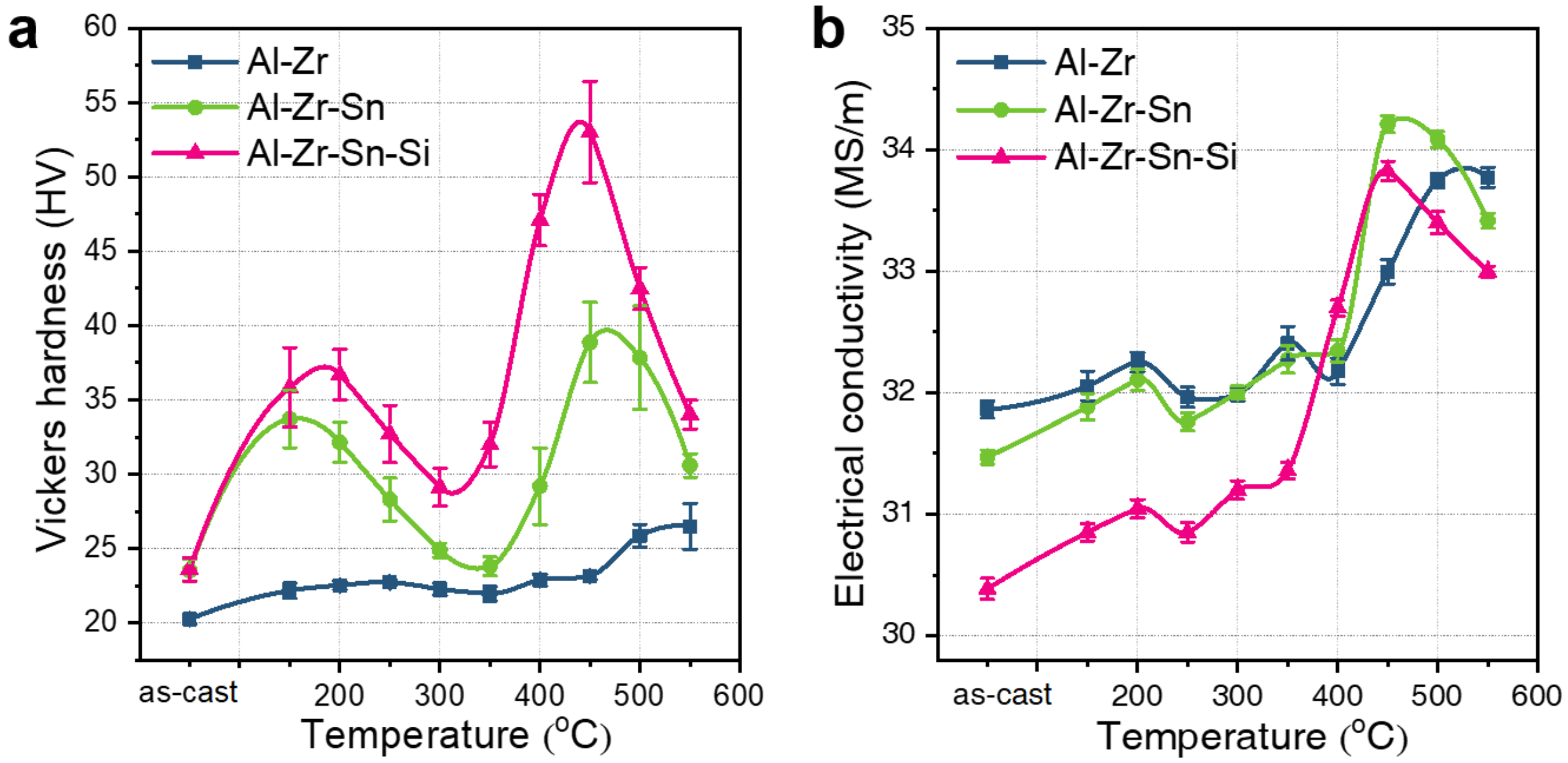

**Supplementary Fig. S3 │Evolution of Vickers microhardness and electrical conductivity of Al-Zr-Sn, Al-Zr-Sn-Si and binary Al-Zr alloys during isochronal aging. a,** Hardness curves. **b,** Electrical conductivity curves. Both two Sn-bearing alloys show typical double-peak in hardness curves, while the binary alloy exhibits a negligible age-hardening effect until heating to 500 °C. The increase in EC is an indication of decomposition of supersaturated solid solution, forming Sn-rich precipitates in low temperature range and Zr-rich precipitates at higher temperatures. A faster increase in EC for Al-Zr-Sn and Al-Zr-Sn-Si alloy in the temperature range of 350~ 450 °C than Al-Zr alloy indicates an enhanced precipitation kinetics of $Al_3Zr$.

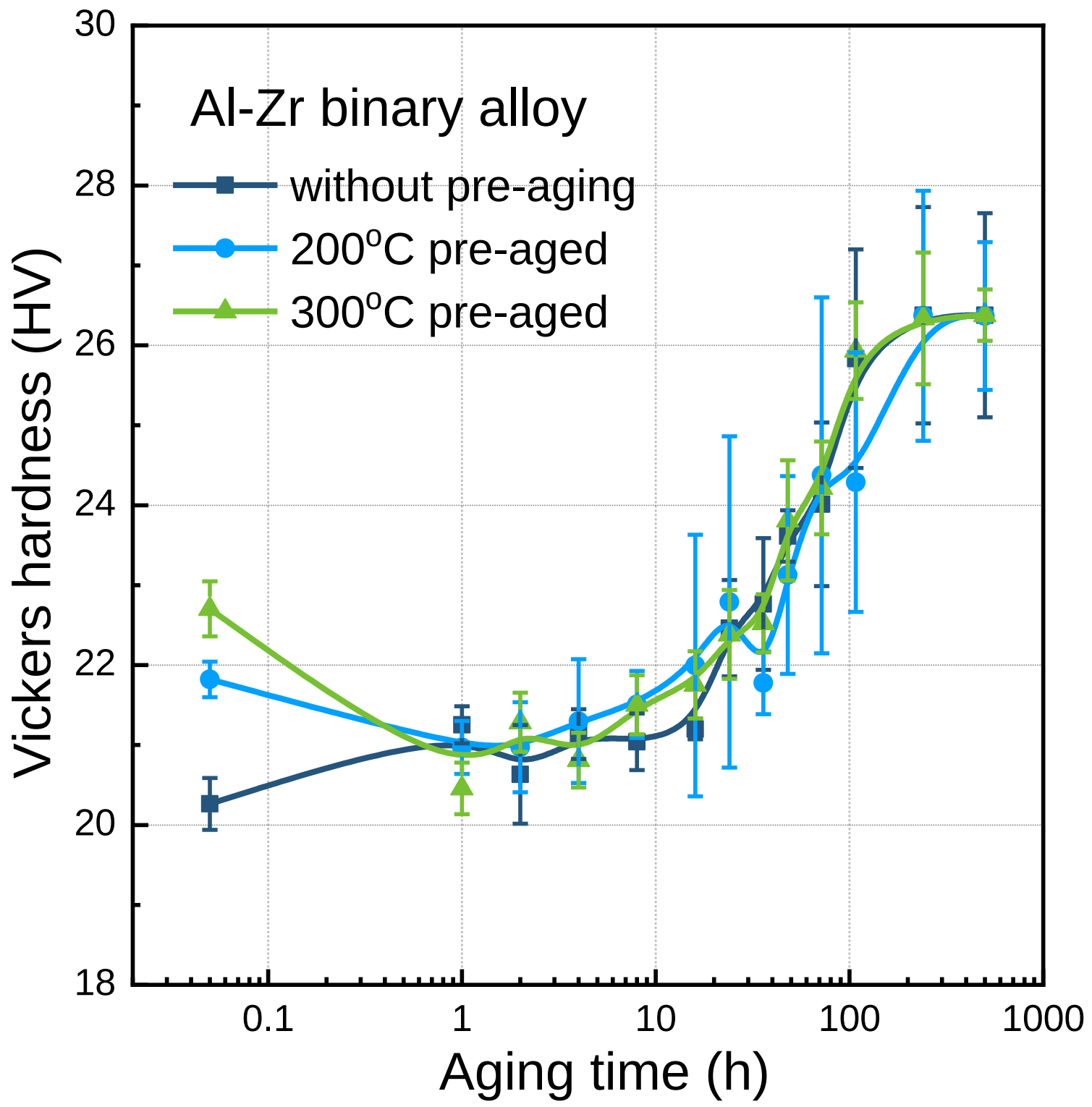

**Supplementary Fig. S4 │Evolution of microhardness of the as-cast, 200 °C/24 h pre-aged and 300 °C/8 h pre-aged samples during isothermal aging at 400 °C, showing preaging has no effect on the precipitation kinetics of $Al_3Zr$ during 400 °C aging.**

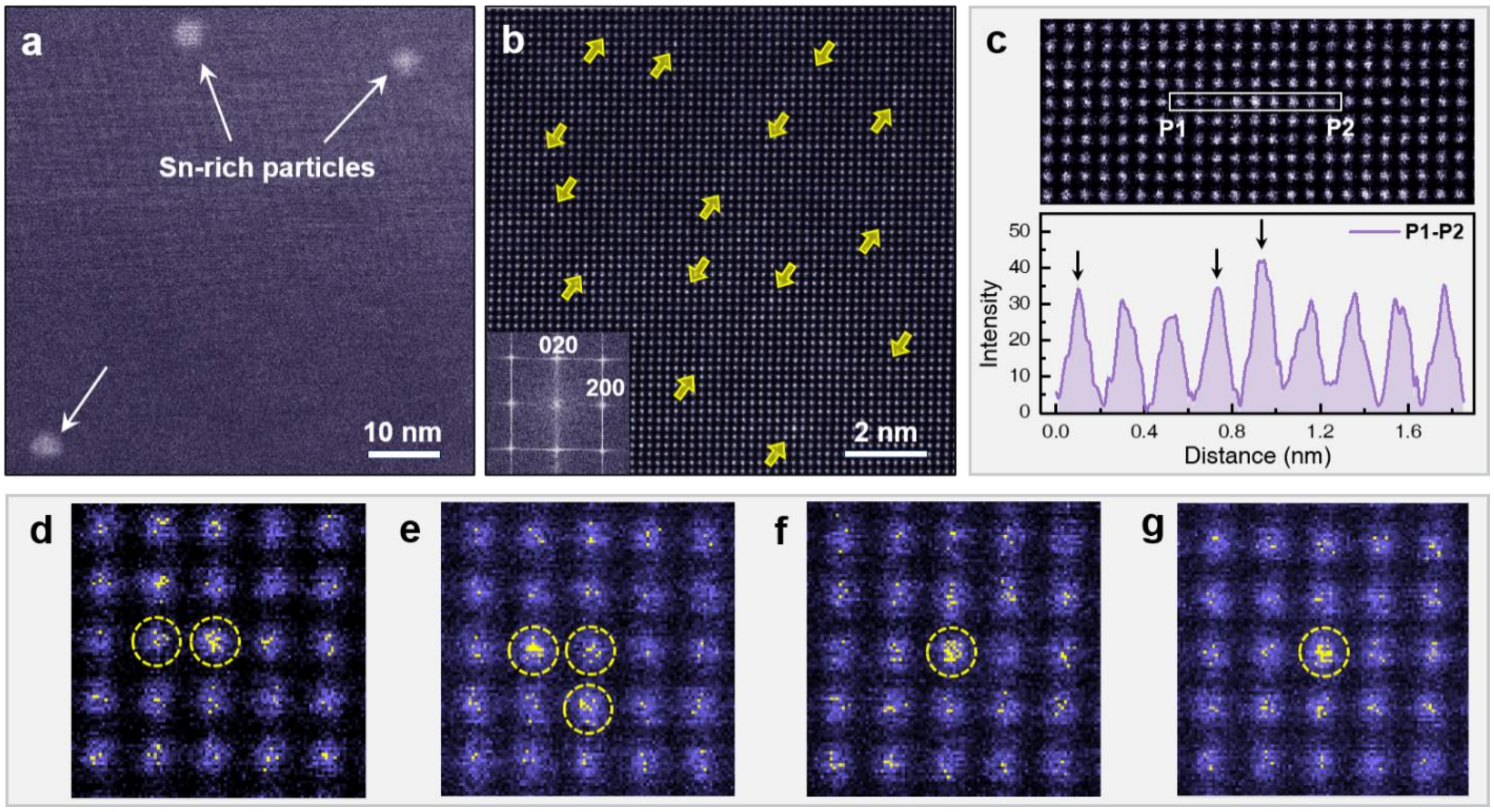

**Supplementary Fig. S5 │ Solute atom complexes forming in Al-Zr-Sn-Si alloy after 200 ºC, 24h pre-aging. a,** Representative HAADF-STEM image showing the presence of fine Sn-rich nanoparticles, which is consistent with the result in Fig. 2b. **b,** Atomic-resolution HAADF-STEM image viewed along <001>$_{Al}$ reveals the tiny solute complexes of heavy atoms as indicated by the yellow arrows in the matrix. The fast Fourier transformation (FFT) pattern in **b** shows only diffraction spots of fcc Al matrix. **c,** Intensity profile of atom columns along P1-P2 line across one possible solute complex. The atom columns with higher intensities as marked by the black arrows are suggested to be enriched with heavy Sn and Zr atoms. **d-g,** Solute atom complexes enlarged from **b**, showing atom columns circled by dashed lines have significantly higher intensity than the surrounding ones.

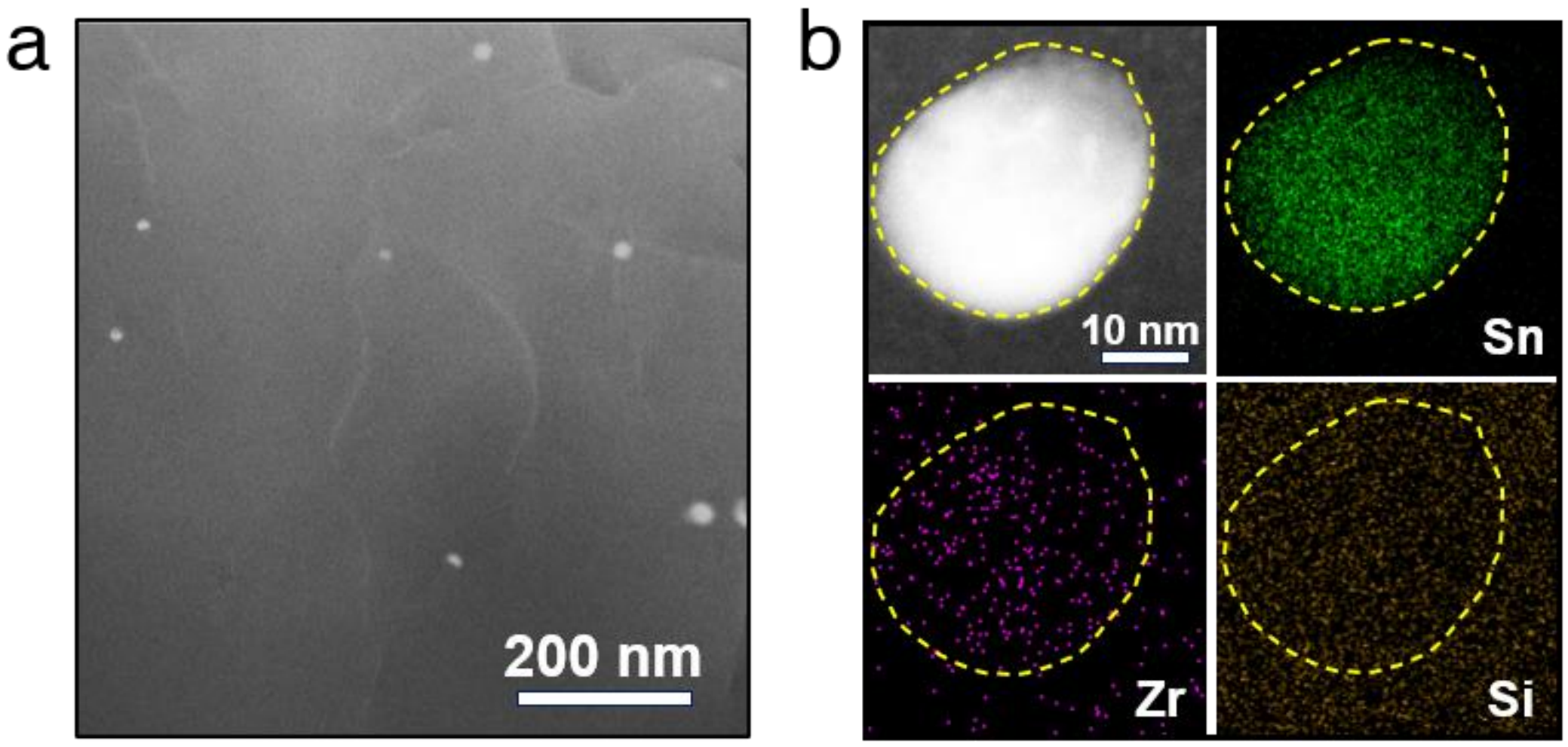

**Supplementary Fig. S6 │ Coarsened Sn-rich particles in Al-Zr-Sn-Si alloy after 300 ºC pre-aging. a,** HAADF-STEM image shows bright elliptic particles with radius of 10-20 nm have formed. **b,** EDX mapping of a typical particle in **a**, showing the particle is a Sn-rich particle with a slight enrichment of Zr.

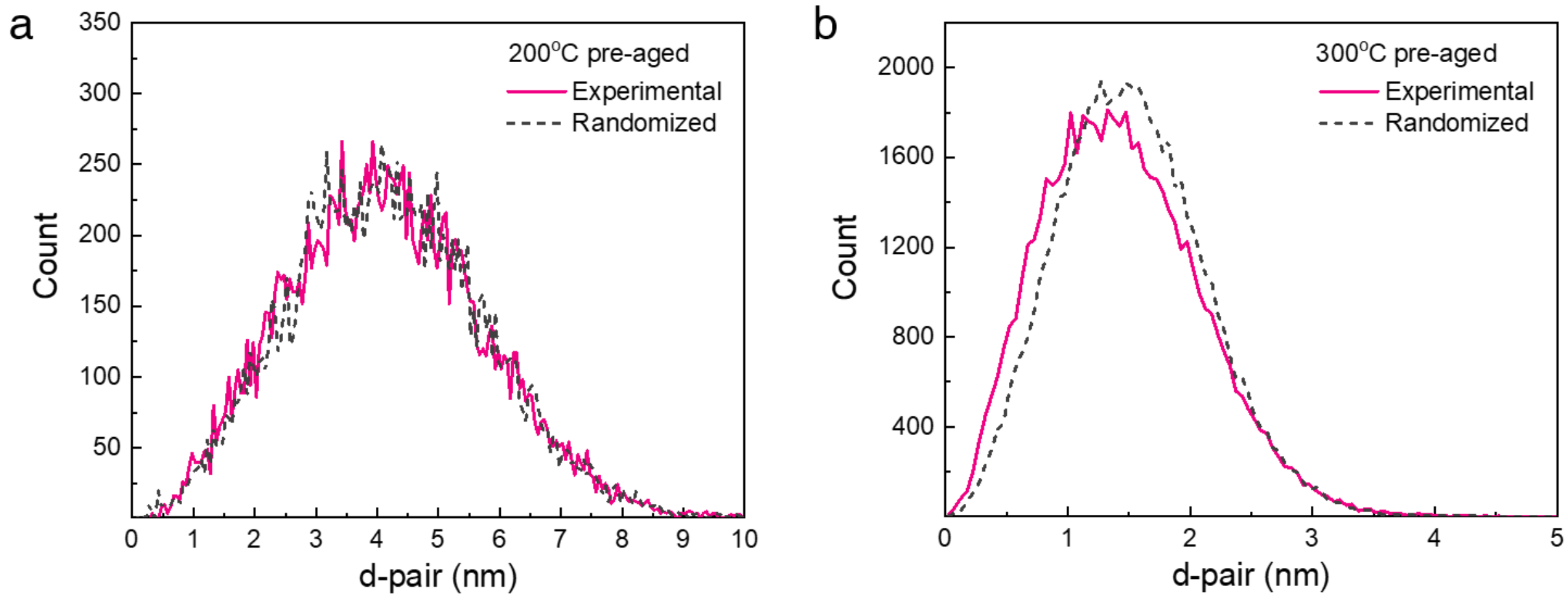

**Supplementary Fig. S7 │The nearest neighbor distribution data of Zr solute atoms in Al-Zr-Sn-Si exported from the APT datasets. a,** 200 °C, 24h pre-aged sample. Zr atoms show a random distribution in the solid solution, indicating the clustering of Zr-Sn atoms is too weak to be distinguished by nearest neighbor distribution method. **b,** 300 °C, 8h pre-aged sample, showing a clear shifting from the random distribution of atoms, confirming a high density of Zr-rich atom clusters/complexes have formed in the solid solution.

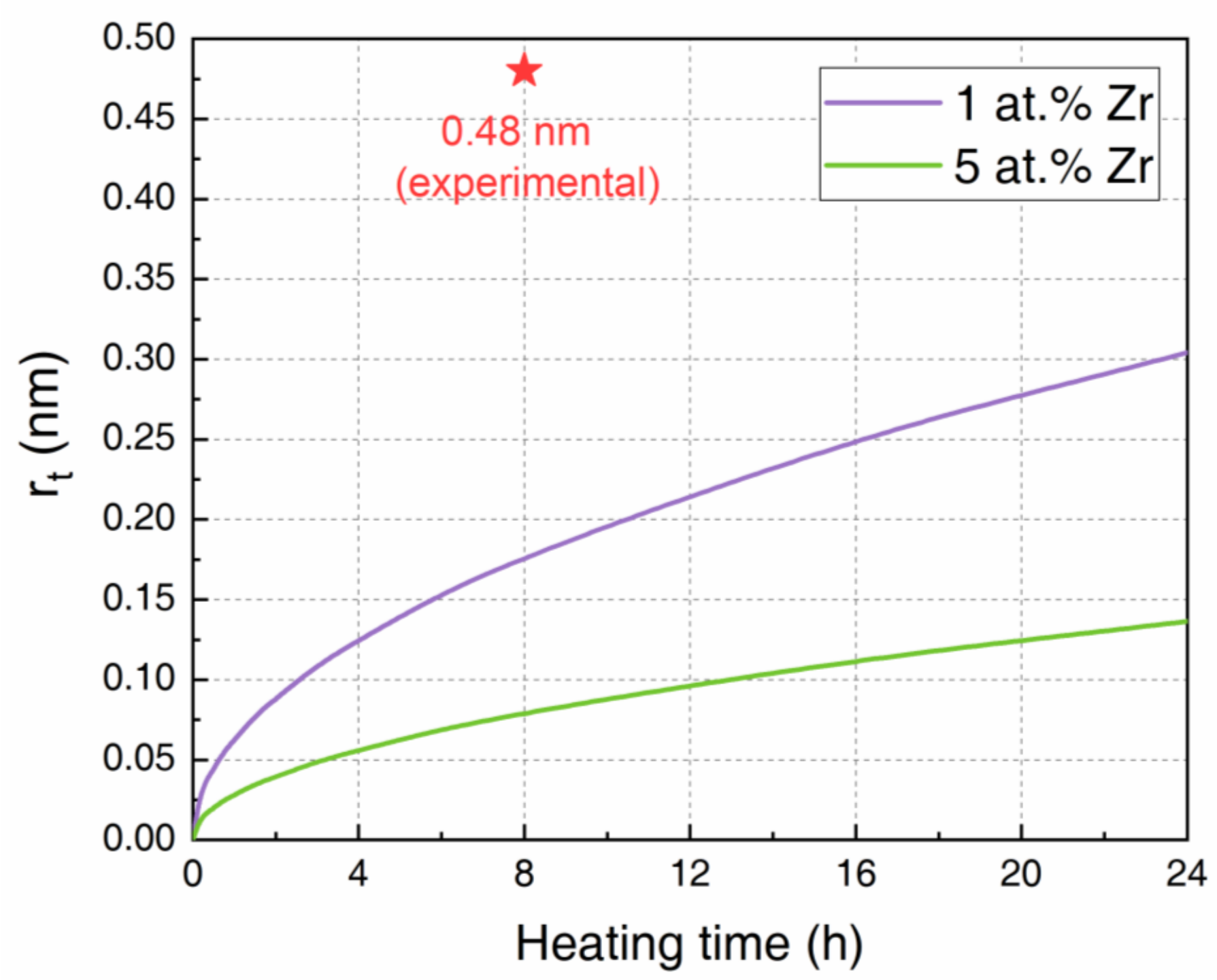

**Supplementary Fig. S8 │Calculated evolution of the radius, $r_t$ , of Zr-rich atom clusters with uniform Zr concentration of 1 at.% and 5 at.%, as a function of heating time during 300 °C pre-aging, in comparison to the experimental data after 300 °C/8h pre-aging (0.48 nm).**

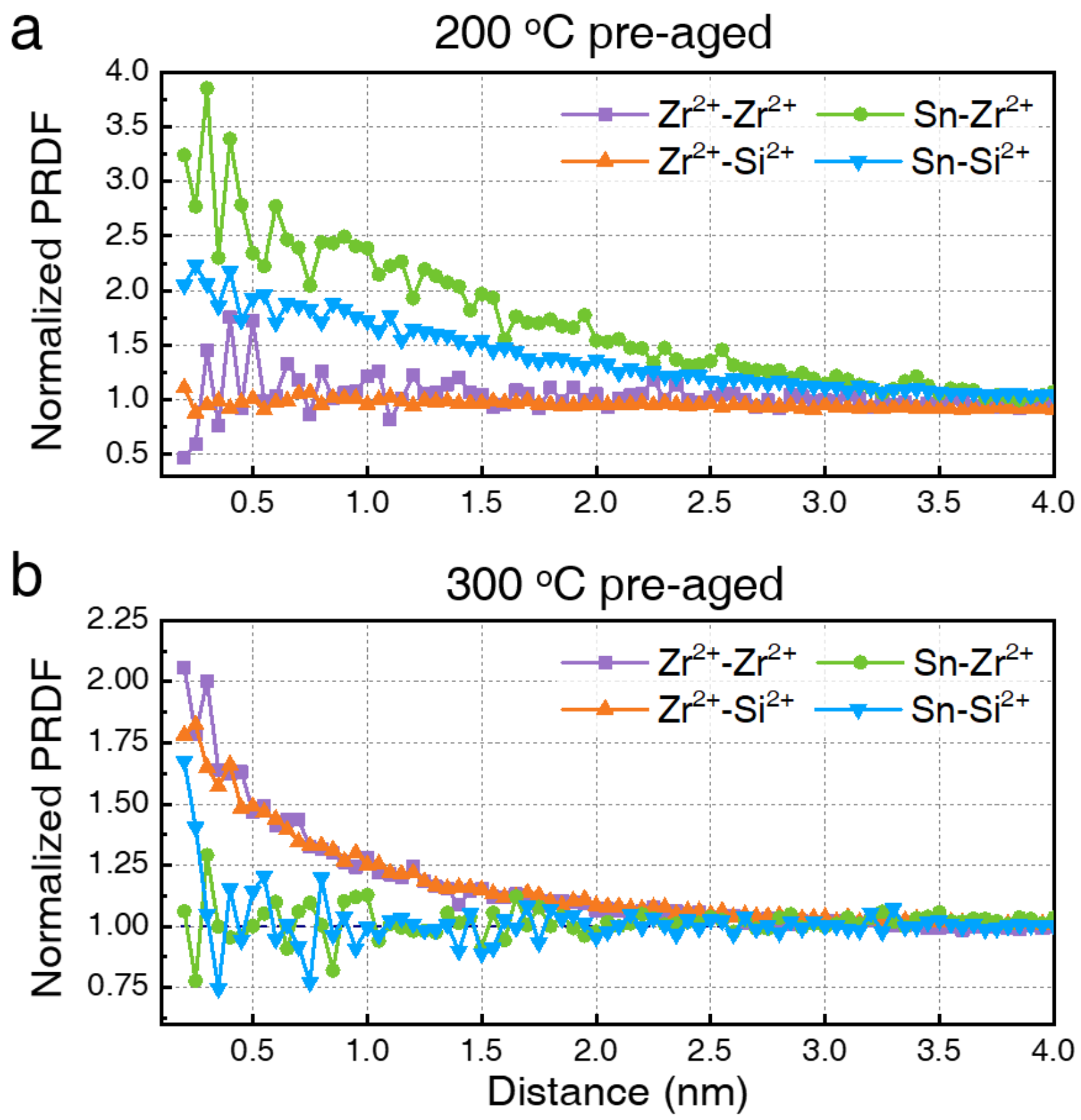

**Supplementary Fig. S9 │Normalized partial radial distribution function (PRDF) of atoms with respect to Zr ($Zr^{2+}$-$Zr^{2+}$, $Zr^{2+}$-$Si^{2+}$) and Sn (Sn-$Zr^{2+}$, Sn-$Si^{2+}$) within the whole reconstructed tips in Fig. 2.** Herein, PRDF data in a distance from 0.2 to 4 nm is provided. **a,** 200 °C, 24h pre-aged sample. A strong Sn-Zr and Sn-Si clustering can be detected within the distance of ~3.5 nm, indicating the precipitation of nano-sized Sn clusters/precipitates with a significant enrichment of Zr and Si atoms. **b,** 300 °C, 8h pre-aged sample. The high normalized PRDF of Zr-Zr and Zr-Si confirms the formation of (Zr, Si)-rich atom clusters, in which a small content of Sn can also be detected. The lower PRDF values of Sn-Zr indicate that the Zr-rich atom clusters have a low concentration of Sn while most of the Sn atoms have formed coarse Sn particles during preaging.

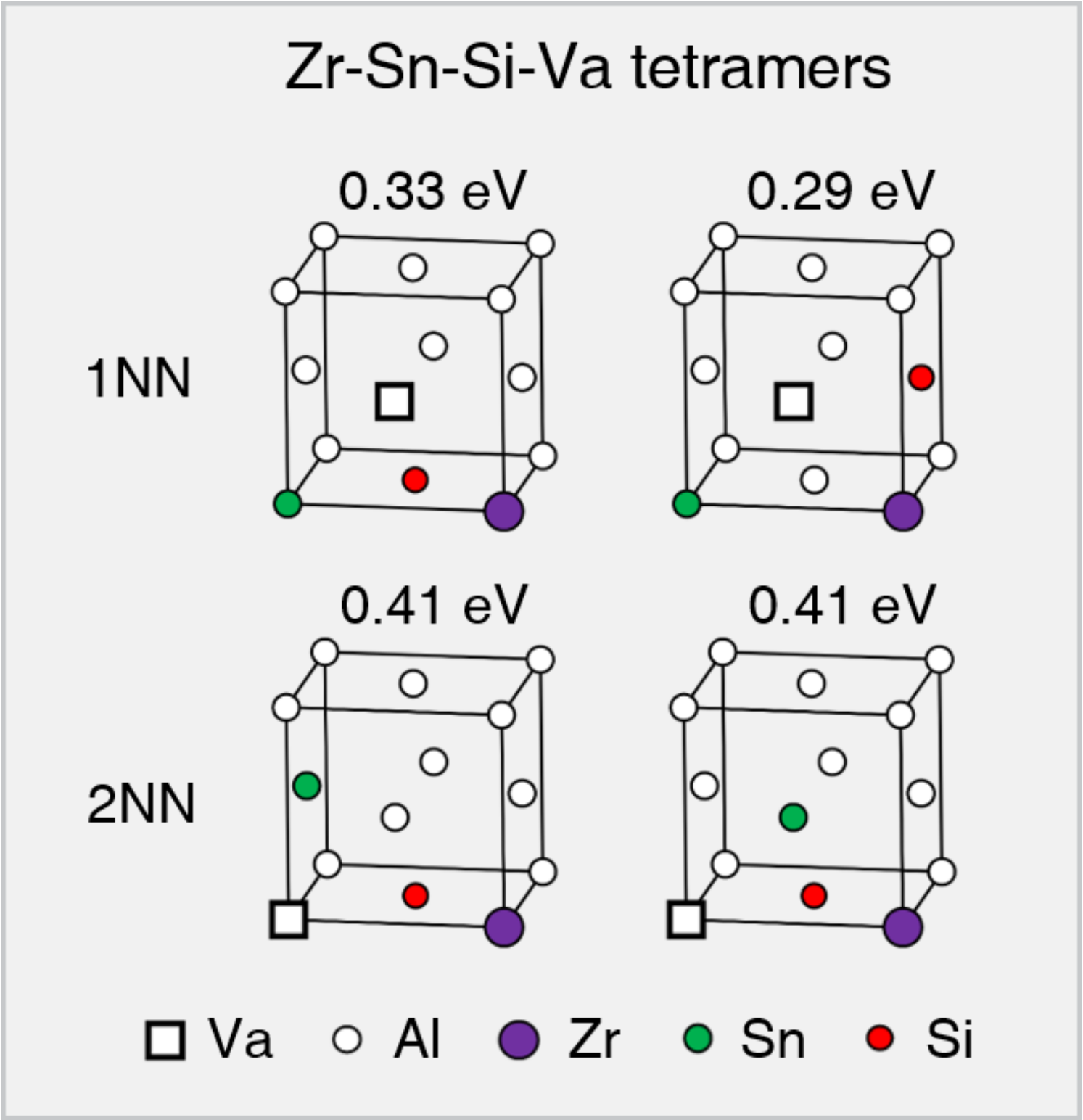

**Supplementary Fig. S10 │Typical atomic configurations of Zr-Sn-Si-Va tetramers with Zr and Va in 1NN and 2NN positions having high binding energies.** As shown in **Supplementary Table S4** and **S5,** 37 different configurations for 1NN and 17 configurations for 2NN of Zr-Sn-Si-Va tetramers were constructed and calculated.

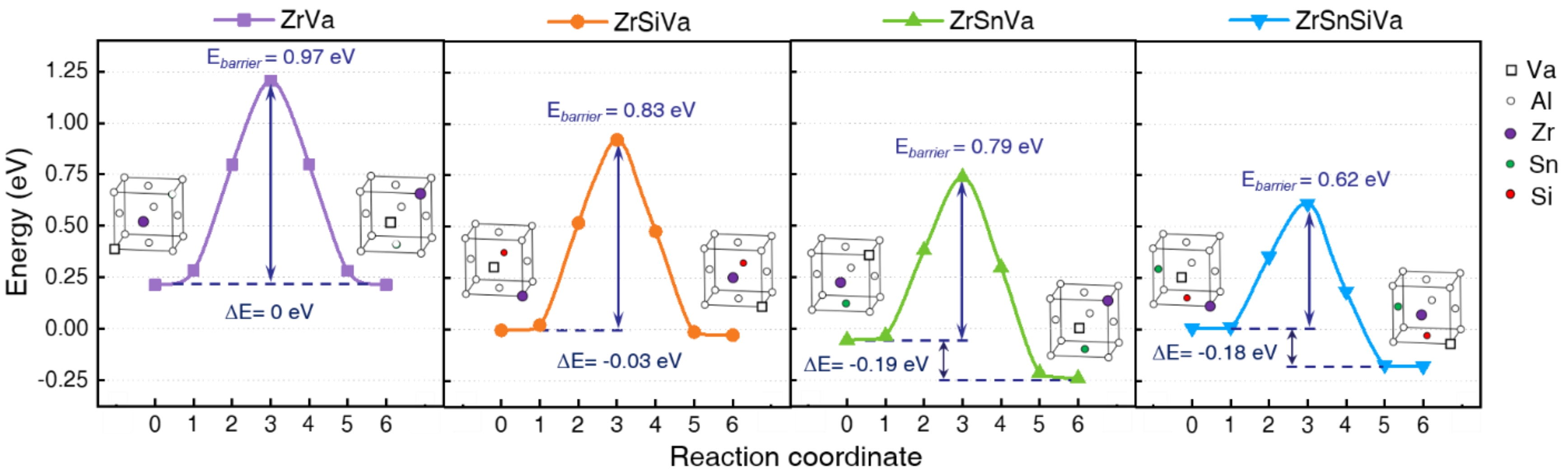

**Supplementary Fig. S11 │Detailed migration energy profile of Zr atoms exchanging position with 1NN vacancy in Al matrix and in different atom complexes shown in Fig. 4b.** For each case, the energy barrier ($E_{barrier}$) of migration, the energy state of initial and final constructions after exchange, and the corresponding total energy change (ΔE) are provided in the figures. With the presence of Si atoms, the formation of Zr-Si-Va trimers can slightly decrease the $E_{barrier}$ from 0.97 eV to 0.83 eV and obtain a reduced ΔE of -0.03 eV. In contrast, Zr-Sn-Va binding exerts much stronger effect in decreasing the $E_{barrier}$ and to 0.79 eV and -0.19 eV, respectively. Zr-Sn-Si-Va tetramers induce the lowest $E_{barrier}$ of 0.62 eV and a considerable reduction in ΔE of -0.18 eV, which provides the most favorable condition for Zr diffusion, enhancing the diffusivity. One should notice that after one-step exchange of Zr and Va, this process can be spontaneously repeated by the fast jump of Sn, Si and Va to form a new Zr-Sn-Si-Va complex with low formation energy, which facilitates the “amoeba” diffusion mechanism.

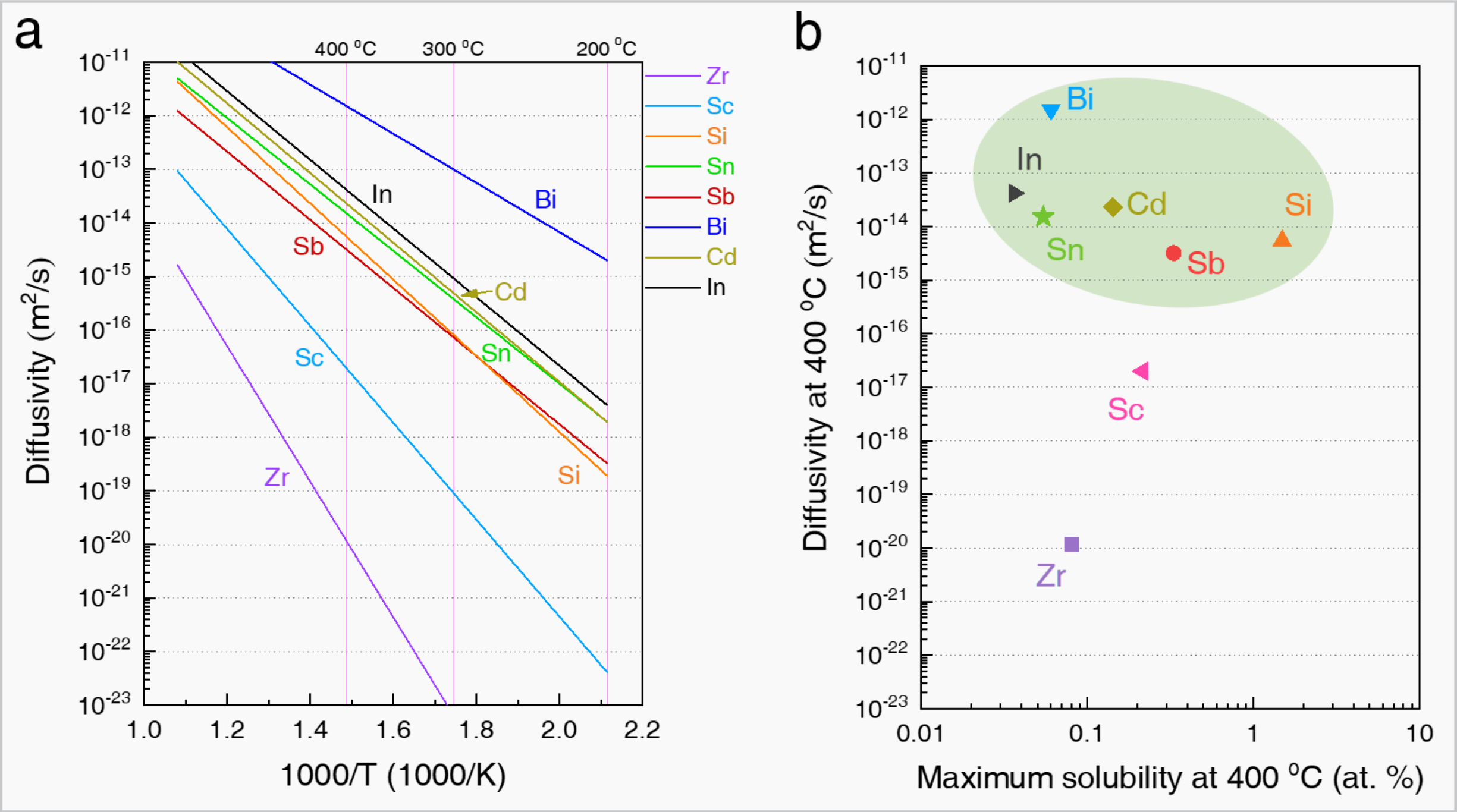

**Supplementary Fig. S12 │Comparison in impurity diffusivities and solubilities of different alloying elements in Al-X binary alloys. a,** Diffusivities of elements contained in the present work as a function of the inverse temperature in Kelvin. The functions are plotted based on the data in Refs [1–4]. **b,** The calculated maximum solubilities of various elements at 400 °C by Thermo-Calc TCAL8 database (www.thermocalc.com) as well as their diffusivities. Note the solubility of Sb is obtained in Ref [5]. It can be found that the dataset of most selected elements, which could promote the $Al_3Zr$ precipitation, resides in the upper-left corner in **b** with much faster diffusivities than Zr and relatively low solubilities in the Al matrix.

**Supplementary Table S1 │ Comparison of the dispersion-hardening efficiency in reported Al-Zr(-RE) series alloys at different peak-aged states from literature data.** Here the dispersion-hardening efficiency is denoted as the equivalent hardness increment (ΔHV) contributed from per 0.01 at. % addition of the $L1_2$-forming elements.

| Alloys composition (at.%) | Heat-treatment regime | $L1_2$ forming element content (at.%) | ΔHV | ΔHV per 0.01at.% | Ref. |
|---|---|---|---|---|---|
| Al–0.07Sc–0.02Yb–0.004Zr | 375 °C isothermal | 0.094 | 19.4 | 2.06 | 6 |
| Al–0.06Sc–0.02Yb–0.02Zr | 375 °C isothermal | 0.1 | 18.15 | 1.815 | |
| Al-0.005Er-0.02Sc-0.07Zr -0.06Si | 400 °C isothermal | 0.095 | 23.4 | 2.46 | 7 |
| | 350 °C /16h+400 °C /8h | 0.095 | 28.55 | 3 | |
| Al-0.005Er-0.02Sc-0.07Zr-0.08V-0.06Si | 400 °C isothermal | 0.095 | 18.86 | 1.98 | |
| | Isochronal | 0.095 | 31.92 | 3.36 | |
| Al-0.10Zr-0.01Sc-0.007Er-0.10Si-0.40Mn-0.09Mo | 400 °C /24 h | 0.117 | 33.04 | 2.82 | 8 |
| Al-0.08Zr-0.014Sc-0.008Er-0.10Si | 400 °C /24 h | 0.102 | 33.1 | 3.24 | |
| Al–0.06Zr | 425 °C/3h isochronal | 0.06 | 9.17 | 1.52 | 9 |
| Al–0.06Sc–0.06Zr | 500 °C /3h isochronal | 0.12 | 38.54 | 3.21 | |
| Al–0.06 Zr–0.06 Sc | 300 °C /24h+400AA | 0.12 | 37.22 | 3.1 | 10 |
| Al–0.06 Zr–0.05 Sc–0.01 Er | 300 °C /24h+400AA | 0.12 | 30.59 | 2.54 | |
| Al–0.06 Zr–0.04 Sc–0.02 Er | 300 °C /24h+400AA | 0.12 | 28.55 | 2.37 | |
| Al–0.1 Zr | 400 °C isothermal | 0.1 | 20.39 | 2.03 | 11 |
| Al–0.1 Zr–0.1 Ti | 400 °C isothermal | 0.2 | 18.86 | 0.94 | |
| Al–0.2 Zr | 400 ºC isothermal | 0.2 | 43.84 | 2.192 | |
| Al–0.07Sc–0.02Yb–0.004Zr | 375 ºC isothermal | 0.094 | 20.39 | 2.16 | 6 |
| Al–0.06Sc–0.02Yb–0.02Zr | 375 ºC isothermal | 0.10 | 21 | 2.1 | |
| Al–0.056Sc–0.007Er–0.021Zr–0.04Si | 400 ºC isothermal/60h | 0.084 | 28.55 | 3.39 | 12 |
| Al-0.085Zr-0.026Si | 400 ºC isothermal | 0.085 | 19.17 | 2.25 | 13 |
| Al-0.14Zr- 0.005Er-0.03Si-0.09 Cr | 400 ºC/260h | 0.145 | 34.16 | 2.35 | |
| Al-0.11Zr- 0.005Er-0.02Si-1Zn | 400 ºC/260h | 0.115 | 33.14 | 2.88 | |
| Al-0.11Zr- 0.005Er-0.02Si | 400 ºC/260h | 0.11 | 32.6 | 2.96 | |
| Al-0.08Zr-0.08Hf-0.045Er-0.03Si-0.04Fe | 350 ºC isothermal | 0.205 | 39.87 | 1.94 | 14 |
| Al-0.1Zr-0.08Hf-0.047Er | 350 ºC isothermal | 0.227 | 28.55 | 1.25 | |
| Al-0.13Zr-0.026Si-0.0045Er-0.076 Ni-0.049Fe | 400 ºC isothermal | 0.1345 | 29.57 | 2.57 | 9 |
| Al-0.12Zr-0.0046Er-0.027Si-0.005Fe | Isochronal | 0.1246 | 33.65 | 2.7 | |
| Al-0.12Zr-0.005Er-0.03W-0.02Si | 400 ºC/100h | 0.125 | 25.49 | 2.03 | 15 |
| Al-0.13Zr-0.005Er-0.02Si | 400 ºC/100h | 0.135 | 26.27 | 1.94 | |
| Al-0.13Zr-0.004Er-0.32Si | 400 ºC/20h | 0.134 | 35.9 | 2.65 | |
| Al-0.12Zr-0.005Er-0.03W-0.38Si | 400 ºC/20h | 0.125 | 35.91 | 2.872 | |
| Al-0.08Zr-0.014Sc-0.008Er-0.10Si | 400 ºC isothermal | 0.102 | 31.61 | 3.09 | 16 |
| | 375 ºC isothermal | 0.102 | 37.01 | 3.62 | |
| Al-0.07Zr | 400 ºC isothermal | 0.07 | 2.6 | 0.37 | |
| Al-0.07Zr-0.007Y | 400 ºC isothermal | 0.077 | 6.5 | 0.84 | 17 |
| Al-0.07Zr-0.019Y | 400 ºC isothermal | 0.089 | 5 | 0.56 | |
| Al–0.04Er–0.04Zr | 350 ºC isothermal | 0.08 | 14.28 | 1.785 | 18 |
| Al–0.04Er–0.08Zr | 350 ºC isothermal | 0.12 | 30.59 | 2.549 | |
| Al–0.03Yb–0.08Zr | 350 ºC isothermal | 0.11 | 15.8 | 1.43 | 19 |
| Al–0.15Si–0.08Zr | 350 ºC isothermal | 0.08 | 9.17 | 1.14 | |
| Al–0.03Yb–0.15Si–0.08Zr | 350 ºC isothermal | 0.11 | 24.06 | 2.18 | |
| | isochronal | 0.11 | 24.47 | 2.22 | |
| Al–0.045Er–0.1Hf–0.08Zr | 375 ºC isothermal | 0.225 | 36.71 | 1.6 | 20 |
| Al–0.06 Zr–0.03 Er | 400 ºC isothermal | 0.09 | 17.64 | 1.96 | 21 |
| Al–0.06 Zr–0.06 Er | 400 ºC isothermal | 0.12 | 19.68 | 1.64 | |
| Al-0.18%Zr-0.19%Fe-0.4%Si | Electromagnetic Casting + rolling + isochronal | 0.18 | 40.79 | 2.26 | 22 |
| Al-0.12%Zr-0.12%Fe-0.023%Si | Aging + HPT + Annealing | 0.12 | 7.8 | 0.65 | 23 |
| Al–0.1Sc–0.1Zr | isochronal | 0.2 | 50.4 | 2.52 | 24 |
| Al-0.06Sc-0.03Zr-0.008Er | 400 ºC /24h | 0.098 | 32.1 | 3.27 | 25 |
| Al-0.06Sc-0.06Zr-0.008Er | 400 ºC /24h | 0.128 | 28 | 2.18 | |
| Al–0.07 Zr-0.065Si-0.06Fe | 400 ºC isothermal | 0.07 | 10.2 | 1.45 | 26 |

**Supplementary Table S2 │Thermodynamic calculation of nucleation of $Al_3Zr$ in Al-Zr binary alloy at different temperatures.**

| T (K) | $C_0$ (at.%) | $C_{Zr}^{eq}$ (at.%) | $\gamma$ (J/m$^2$) | $D_{Zr}$ (m$^2$/s) | $\Delta G_v$ (J/m$^3$) | $\Delta G_\varepsilon$ (J/ m$^3$) | $\Delta G^*$ (J/ m$^3$) | $\Delta G^*/k_bT$ | $r^0$ (nm) | $\tau$(s) | $\tau$(s) |
|---|---|---|---|---|---|---|---|---|---|---|---|
| 473 | 0,0801 | 0.00015 | 0.1 | $1.34\times 10^{-28}$ | $-6.17\times 10^{8}$ | $2.37\times 10^{7}$ | $4.79\times 10^{-20}$ | 7.3 | 0.34 | $2.25\times 10^{12}$ | $9.56\times 10^{11}$ |
| 573 | 0,0801 | 0.00214 | 0.1 | $6.20\times 10^{-24}$ | $-4.32\times 10^{8}$ | $2.37\times 10^{7}$ | $1.02\times 10^{-19}$ | 12.8 | 0.49 | $1.16\times 10^{8}$ | $4.92\times 10^{7}$ |
| 673 | 0,0801 | 0.0138 | 0.1 | $1.18\times 10^{-20}$ | $-2.46\times 10^{8}$ | $2.37\times 10^{7}$ | $3.44\times 10^{-19}$ | 37.1 | 0.91 | $2.00\times 10^{5}$ | $8.49\times 10^{4}$ |
| 723 | 0,0801 | 0.0289 | 0.1 | $2.35\times 10^{-19}$ | $-1.53\times 10^{8}$ | $2.37\times 10^{7}$ | $1.02\times 10^{-18}$ | 103 | 1.56 | $2.40\times 10^{4}$ | $1.02\times 10^{4}$ |

**Supplementary Table S3 | Calculated energies of formation (eV/atom) of solute elements X = Sn, Sb, Bi and Si substituting into the lattices of $Al_3Zr$ and bulk Al.**

| Substituting sites | Sn | Sb | Bi | Si |
|---|---|---|---|---|
| Al sublattice in bulk $Al_3Zr$ | 0.23 | 0.57 | 0.85 | 0.12 |
| Zr sublattice in bulk $Al_3Zr$ | 0.66 | 0.85 | 1.02 | 0.20 |

**Supplementary Table S4 │ Calculated binding energies (eV) in 1NN for Zr-Sn-Si-Va tetramers with 37 different configurations within the Al supercell.** Herein, 1NN refers to the relative location for Zr-Va. Schematics of two atomic configurations with the 1st and 2nd maximum binding energy values are displayed in **Fig. S9**.

| Configurations | Binding energy (eV) | Detailed NN relationships between two constituents | | | | | |
|---|---|---|---|---|---|---|---|
| | | Zr-Va | Zr-Si | Zr-Sn | Sn-Si | Sn-Va | Si-Va |
| 1 | 0.11 | 1 | 2 | 2 | 4 | 1 | 1 |
| 2 | 0.32 | 1 | 1 | 2 | 1 | 1 | 1 |
| 3 | 0.17 | 1 | 3 | 2 | 3 | 1 | 1 |
| 4 | 0.29 | 1 | 1 | 2 | 3 | 1 | 1 |
| 5 | 0.16 | 1 | 3 | 2 | 1 | 1 | 1 |
| 6 | 0.18 | 1 | 1 | 2 | 5 | 1 | 2 |
| 7 | 0.26 | 1 | 1 | 2 | 1 | 1 | 2 |
| 8 | 0.19 | 1 | 1 | 4 | 3 | 1 | 1 |
| 9 | 0.09 | 1 | 1 | 4 | 5 | 1 | 2 |
| 10 | 0.08 | 1 | 2 | 1 | 1 | 1 | 1 |
| 11 | 0.02 | 1 | 2 | 1 | 3 | 1 | 1 |
| 12 | 0.03 | 1 | 4 | 1 | 3 | 1 | 1 |
| 13 | 0.15 | 1 | 1 | 1 | 2 | 1 | 1 |
| 14 | 0.03 | 1 | 3 | 1 | 2 | 1 | 1 |
| 15 | 0.05 | 1 | 3 | 1 | 4 | 1 | 1 |
| 16 | 0.14 | 1 | 1 | 1 | 1 | 1 | 1 |
| 17 | 0.18 | 1 | 1 | 1 | 3 | 1 | 1 |
| 18 | 0.05 | 1 | 3 | 1 | 1 | 1 | 1 |
| 19 | 0.05 | 1 | 3 | 1 | 3 | 1 | 1 |
| 20 | 0.00 | 1 | 3 | 1 | 1 | 1 | 2 |
| 21 | 0.01 | 1 | 3 | 1 | 5 | 1 | 2 |
| 22 | 0.10 | 1 | 1 | 1 | 3 | 1 | 2 |
| 23 | 0.00 | 1 | 5 | 1 | 3 | 1 | 2 |
| 24 | 0.07 | 1 | 1 | 1 | 1 | 1 | 2 |
| 25 | 0.00 | 1 | 5 | 1 | 5 | 1 | 2 |
| 26 | 0.01 | 1 | 2 | 3 | 3 | 1 | 1 |
| 27 | 0.02 | 1 | 2 | 3 | 1 | 1 | 1 |
| 28 | 0.03 | 1 | 4 | 3 | 1 | 1 | 1 |
| 29 | 0.17 | 1 | 1 | 3 | 2 | 1 | 1 |
| 30 | 0.18 | 1 | 1 | 3 | 4 | 1 | 1 |
| 31 | 0.03 | 1 | 3 | 3 | 2 | 1 | 1 |
| 32 | 0.18 | 1 | 1 | 3 | 1 | 1 | 1 |
| 33 | 0.18 | 1 | 1 | 3 | 3 | 1 | 1 |
| 34 | 0.05 | 1 | 3 | 3 | 1 | 1 | 1 |
| 35 | 0.05 | 1 | 3 | 3 | 3 | 1 | 1 |
| 36 | 0.08 | 1 | 1 | 3 | 3 | 1 | 2 |
| 37 | 0.08 | 1 | 1 | 3 | 5 | 1 | 2 |

**Supplementary Table S5 | Calculated binding energies (eV) in 2NN for Zr-Sn-Si-Va tetramers with 17 different configurations within the Al supercell.** Herein, 2NN refers to the relative location for Zr-Va. Schematics of two atomic configurations with the 1st and 2nd maximum binding energy values are displayed in Fig. 4B.

| Configurations | Binding energy (eV) | Detailed NN relationships between two constituents | | | | | |
|---|---|---|---|---|---|---|---|
| | | Zr-Va | Zr-Si | Zr-Sn | Sn-Si | Sn-Va | Si-Va |
| 1 | 0.35 | 2 | 3 | 3 | 2 | 1 | 1 |
| 2 | 0.37 | 2 | 3 | 3 | 4 | 1 | 1 |
| 3 | 0.41 | 2 | 1 | 3 | 1 | 1 | 1 |
| 4 | 0.34 | 2 | 5 | 3 | 1 | 1 | 1 |
| 5 | 0.39 | 2 | 1 | 3 | 3 | 1 | 1 |
| 6 | 0.33 | 2 | 5 | 3 | 3 | 1 | 1 |
| 7 | 0.33 | 2 | 5 | 1 | 2 | 1 | 1 |
| 8 | 0.41 | 2 | 1 | 1 | 2 | 1 | 1 |
| 9 | 0.35 | 2 | 5 | 1 | 4 | 1 | 1 |
| 10 | 0.36 | 2 | 3 | 1 | 1 | 1 | 1 |
| 11 | 0.38 | 2 | 3 | 1 | 3 | 1 | 1 |
| 12 | 0.41 | 2 | 1 | 1 | 1 | 1 | 1 |
| 13 | 0.36 | 2 | 5 | 1 | 3 | 1 | 1 |
| 14 | 0.30 | 2 | 8 | 1 | 5 | 1 | 2 |
| 15 | 0.36 | 2 | 1 | 5 | 2 | 1 | 1 |
| 16 | 0.38 | 2 | 1 | 5 | 4 | 1 | 1 |
| 17 | 0.38 | 2 | 1 | 5 | 3 | 1 | 1 |